\documentclass[journal,10pt]{IEEEtran}
\makeatletter
\def\endthebibliography{% 
    \def\@noitemerr{\@latex@warning{Empty `thebibliography' environment}}%
    \endlist
}
\makeatother

\usepackage{cite}

\usepackage[pdftex]{graphicx}
\usepackage[caption=false,font=footnotesize]{subfig}

\usepackage{amsmath}
\usepackage{mathtools, cuted}
\usepackage{amssymb}
\usepackage{bm}
\usepackage{mathrsfs}
\usepackage{breqn}
\usepackage{braket}
 
\usepackage{pifont}
\usepackage{xcolor}
\usepackage{url}
\usepackage{lettrine}
\usepackage{lipsum}
\usepackage{siunitx}
\usepackage{soul}
\usepackage{array}
\usepackage[inline]{enumitem}
\usepackage{epsfig}
\usepackage{filecontents}
\usepackage{booktabs}

\usepackage{algpseudocode}
\usepackage{algorithm, tabularx}
\usepackage{multirow}
\newcolumntype{L}[1]{>{\raggedright\let\newline\\\arraybackslash\hspace{0pt}}m{#1}}
\newcolumntype{C}[1]{>{\centering\let\newline\\\arraybackslash\hspace{0pt}}m{#1}}
\newcolumntype{R}[1]{>{\raggedleft\let\newline\\\arraybackslash\hspace{0pt}}m{#1}}
\newlength{\maxwidth}

\makeatletter
\newcommand{\multiline}[1]{%
	\begin{tabularx}{\dimexpr\linewidth-\ALG@thistlm}[t]{@{}X@{}}
		#1
	\end{tabularx}
}
\makeatother
\algdef{SE}[SUBALG]{Indent}{EndIndent}{}{{\algorithmicend\ }}
\algtext*{Indent}
\algtext*{EndIndent}

\usepackage{amsthm}

\theoremstyle{remark}

\DeclareMathOperator*{\argmin}{\arg\min}
\DeclareMathOperator*{\argmax}{\arg\max}

\newcommand{\cmark}{\ding{51}}%
\begin{document}

\title{Quantum-PROBE: Rydberg Atomic Receiver-Based Multi-AoA Estimation with RF Lens}

\author{Hong-Bae Jeon,~\IEEEmembership{Member,~IEEE,} Kaibin Huang,~\IEEEmembership{Fellow,~IEEE}, and Chan-Byoung Chae,~\IEEEmembership{Fellow,~IEEE}%
\thanks{This work was supported by Hankuk University of Foreign Studies Research Fund of 2026. \textit{(Corresponding: Chan-Byoung Chae.)}}%
\thanks{H.-B. Jeon is with the Department of Information Communications Engineering, Hankuk University of Foreign Studies, Yong-in, 17035, Korea (e-mail: hongbae08@hufs.ac.kr).}%
\thanks{K. Huang is with the Department of Electrical and Electronic Engineering, The University of Hong Kong, Hong Kong (e-mail: huangkb@eee.hku.hk).}
\thanks{C.-B. Chae is with the School of Integrated Technology, Yonsei University, Seoul 03722, Korea (e-mail: cbchae@yonsei.ac.kr).}%
}

\maketitle

\begin{abstract}
This paper presents the Quantum-Power pROfile Based Estimation (PROBE) framework, a Rydberg Atomic Receiver (RARE)-based multi-user angle-of-arrival (AoA) estimation approach equipped with a radio-frequency (RF) lens front end. We establish a physics-consistent analytical model showing that magnitude-only RARE measurements, processed via the beam-propagation method (BPM) and snapshot-wise power accumulation, can be rigorously characterized as a nonnegative superposition of AoA-dependent, lens-induced spatial power profiles. This formulation reveals a structured and interpretable power-domain dictionary that enables multi-user AoA recovery without explicit phase reconstruction. Building on this foundation, we develop two complementary recovery strategies: (i) a principled non-negative least absolute shrinkage and selection operator (NN-LASSO)-based solver that estimates a sparse nonnegative angular representation via an accelerated proximal-gradient method followed by cluster-based AoA decoding, and (ii) a low-complexity successive interference cancellation (SIC) algorithm that iteratively identifies and removes dominant power-profile components through cosine-similarity matching. Simulation results demonstrate that the proposed Quantum-PROBE framework consistently outperforms representative RARE- and RF-based benchmarks across diverse system configurations, while offering a clear accuracy-complexity tradeoff between the NN-LASSO and SIC variants for practical quantum sensing deployments.

\end{abstract}


\IEEEpeerreviewmaketitle

\section{Introduction}
\IEEEPARstart{R}{ecently}, Rydberg Atomic REceiver (RARE)~\cite{atomicmag, atomicjsac, rarnature} have gained increasing attention as an enabling technology for sixth-generation (6G) wireless network, by exploiting the quantum-mechanical properties of highly excited Rydberg atoms for radio-frequency (RF) electric-field detection. A Rydberg atom is formed when one or more electrons are excited from the ground state to a high-lying energy level, giving rise to pronounced quantum responses to incident RF fields. Building upon this property, RARE utilizes phenomena such as electromagnetically induced transparency (EIT) and Autler-Townes (AT) splitting to transduce RF field variations into measurable optical signatures, thereby enabling reliable recovery of the transmitted information~\cite{tqe}. With the maturation of this quantum sensing paradigm, RARE has experimentally demonstrated the capability to extract a wide range of RF signal attributes, including amplitude~\cite{rydamp}, phase~\cite{rarclose2}, and polarization~\cite{polarization_aware_Doa}, with unprecedented sensitivity and precision~\cite{RAR_Classical_Comm_Sensing}.

Beyond extending the functionality of classical receivers, RARE fundamentally departs from antenna-based RF front-ends by circumventing the thermal-noise bottleneck inherent to metal conductors, since the atom-field interaction itself does not generate thermal noise~\cite{atomicjsac, Harnessing_RAR_Tutorial}. Moreover, the quantum shot noise (QSN) associated with probing Rydberg states is typically several orders of magnitude lower than the conventional thermal noise floor~\cite{quanmobi}, enabling reception close to the standard quantum limit (e.g., order of $\mathrm{nVcm^{-1}Hz^{-1/2}}$~\cite{qsens, rarclose3}). Indeed, it is reported that the noise-equivalent power (NEP) of RARE can be typically 20-30~dB less than that of the most sensitive traditional wireless receivers~\cite{rqs}.

Furthermore, for conventional conductor-based antennas, effective interaction with an incident electromagnetic (EM) wave requires the antenna length to be on the order of the carrier wavelength, so that the induced current can oscillate efficiently along the structure~\cite{antenna5g}. As a consequence, the operational frequency bands of classical antennas are inherently constrained by their physical dimensions, making wideband or multi-band reconfiguration challenging without hardware modifications. In sharp contrast, EM waves do not need to induce resonant currents on RARE for signal detection~\cite{efm, rarclose1}. Instead, RARE detects incident radiation through photon-atom interactions enabled by the wave-particle duality of EM waves~\cite{foot, rarclose2}. Owing to this fundamentally different reception mechanism, RARE can be flexibly tuned across ultra-wide frequency ranges by selecting appropriate atomic energy-level transitions, without altering the underlying hardware structure~\cite{efm, rydhard1, rydhard2}. 
\begin{table*}[t]
\centering
\caption{Comparison of RARE-Based AoA Estimation Methods}
\label{tab:comparison}
\renewcommand{\arraystretch}{1.2}
\begin{tabular}{lccccc}
\toprule
\textbf{Reference} & \cite{qmusic} & \cite{wsat} & \cite{TCOM25_Single_RAR_AoA} & \cite{raoa} & \textbf{This Work} \\
\midrule
Multi-User AoA Estimation Capability & \cmark & \cmark & -- & -- & \cmark \\
Exploitation of Multiple RARE Elements    & \cmark & \cmark & -- & -- & \cmark \\
Low-Complexity Scalability (Linear in the Number of RARE Elements) & -- & -- & \cmark & -- & \cmark \\
Phase-Free AoA Estimation Based on Power Profiles    & -- & -- & -- & -- & \cmark \\
\bottomrule
\end{tabular}
\end{table*}

These distinctive advantages position RARE as compelling candidates for next-generation wireless reception under extremely weak electromagnetic fields, such as satellite and space-air-ground communication links where ultra-high sensitivity is essential~\cite{antsp, tqe}, thereby surpassing the inherent limitations of conventional antenna-based receivers. Motivated by these merits, a growing body of research has actively explored RARE-based architectures along a clear evolutionary trajectory, spanning from physical-layer signal recovery to system-level integration. Early studies primarily focused on validating the feasibility of modulation and demodulation through quantum sensing mechanisms. In~\cite{atomicjsac}, the authors demonstrated the recovery of amplitude- and frequency-modulated (AM/FM) signals using RARE and subsequently extended this framework to the multi-user regime by modeling atomic single/multiple-input-multiple-output (SIMO/MIMO) reception as a biased phase-retrieval problem, together with an expectation-maximization Gerchberg-Saxton (EM-GS) algorithm for jointly decoding multi-user symbols. From the transmission perspective,~\cite{Precoding_atomicMIMO} established an atomic-MIMO communication model and revealed its fundamental departure from conventional RF MIMO systems, characterized by a nonlinear magnitude-only input-output relationship. To address this challenge, an In-phase-and-
Quadrature (IQ)-aware precoding strategy was proposed, which was shown to achieve the capacity limit of atomic MIMO systems. Building upon these physical-layer insights, more recent efforts have shifted toward system-level deployment and scalability. For instance,~\cite{RAR_MU_MIMO_Uplink} demonstrated that a single RARE front-end can jointly demodulate spatially multiplexed uplink signals through quantum-optical processing, thereby validating the feasibility of multi-user connectivity. In parallel,~\cite{RAQ_MIMO_Multiband} proposed a Rydberg atomic quantum-MIMO (RAQ-MIMO) architecture, showing that a vapor-cell-based array can simultaneously process multiple RF bands by leveraging quantum transconductance modeling and weighted minimum mean-square error (WMMSE) optimization. Together, these studies highlight the rapid progression of RARE research from proof-of-concept signal recovery toward practical, multi-user, and multi-band quantum wireless systems.Collectively, these results suggest that RARE is rapidly emerging as a compelling candidate for 6G receiver front-ends.

Along with the rapid progress in quantum-enhanced communication using RARE, increasing attention has been directed toward exploiting their capability for spatial information extraction in wireless sensing, which can jointly leveraging communication and sensing functionalities within a unified RARE-based framework~\cite{New_Paradigm_RAR_ISAC}. In particular, angle-of-arrival (AoA) estimation plays a pivotal role in wireless sensing, as it constitutes a fundamental building block for localization, tracking, interference awareness, and spatial resource management in emerging 6G applications~\cite{spmdoa, jsacdoa}. Despite its importance, AoA estimation in RARE-based systems remains in an early stage and poses unique challenges. Unlike conventional antenna arrays that provide phase-coherent observations, RARE inherently yields magnitude-only measurements governed by quantum sensing mechanisms~\cite{Precoding_atomicMIMO}, rendering many classical subspace-based or phase-dependent AoA estimation techniques inapplicable.

To address this issue,~\cite{qmusic} introduced Quantum multiple signal classification (MUSIC) for multi-user AoA estimation in RARE systems, while~\cite{wsat} extended RARE-based sensing to multi-band scenarios by jointly estimating AoAs across different carrier frequencies, both considering the phase recovery by modifying the EM-GS Algorithm. More recently,~\cite{TCOM25_Single_RAR_AoA} demonstrated that even a single RARE can infer AoA by exploiting inner-vapor interference effects within the atomic medium, further highlighting the spatial sensing potential of RARE. In addition, the authors in~\cite{raoa} utilized simple geometric relationships between phase differences and AoA for single-target estimation. However, existing approaches still suffer from several fundamental limitations. In particular,~\cite{qmusic, wsat} rely on intermediate phase recovery, which introduces additional computational complexity and error propagation under magnitude-only atomic observations. Moreover,~\cite{TCOM25_Single_RAR_AoA, raoa} do not exploit the spatial diversity offered by multiple RARE elements and are inherently limited to single-target scenarios. To the best of our knowledge, there is still no AoA estimation framework with low-complexity that directly leverages the RARE power-domain characteristics to enable reliable multi-user AoA estimation utilizing multiple RARE elements without explicit phase reconstruction, which motivates the development of a new AoA estimation paradigm.

Focusing on the power profile induced by RARE-based reception, a complementary line of research in classical RF systems has demonstrated the utility of RF lenses for spatial signal processing by virtue of their inherent power-focusing properties~\cite{lenscsmag, lensntn, zeng2016millimeterWM}. An RF lens, a dielectric or engineered structure placed in front of an antenna array, bends and concentrates incident electromagnetic waves in a manner analogous to an optical lens, thereby converting angular information into spatially localized power distributions across the array aperture~\cite{lensmeta}. This focusing effect has been widely exploited for, e.g., millimeter-wave (mmWave) MIMO communications~\cite{lensmm, leaklens} or non-terrestrial network (NTN)~\cite{lensntn, antsp}, where it enables high directivity and beamforming gains without the need for complex phase-shifter networks, significantly enhancing link performance and system capacity~\cite{cholens, Squint, trimimo, trimimo22, kwon2016rf_LE}.

Beyond communications, RF lens architectures have been extensively investigated for sensing and localization applications by exploiting their inherent power-focusing property. Early theoretical studies, such as~\cite{CRLB}, analytically characterized the fundamental limits of AoA estimation in RF lens antenna arrays by deriving the Cramér-Rao bound (CRB), explicitly revealing that the lens-induced non-uniform power profile provides intrinsic angular discriminability gains over conventional uniform arrays. Building upon this insight, subsequent works demonstrated how such power focusing can be practically exploited for AoA estimation. In~\cite{twcdoa}, the authors showed that an RF lens effectively transforms AoA information into spatially concentrated power distributions across the antenna array, enabling accurate AoA estimation by directly operating in the power domain without relying on phase-coherent processing. Similarly,~\cite{jsacdoa} demonstrated that lens-based MIMO systems can significantly simplify AoA estimation and localization by physically separating signals impinging from different directions, thereby improving robustness against interference and noise. To further reduce implementation complexity,~\cite{suk} proposed a low-complexity AoA estimation algorithm that leverages the distinctive lens-induced power profile, showing that reliable angular estimation can be achieved with substantially reduced computational overhead. More recently, the spatial power-focusing capability of RF lenses has been extended to practical system-level applications, such as one-shot coarse pointing in hybrid RF/free-space optical (FSO) communications~\cite{HJFSO}, where the focal position of the received power directly provides reliable directional information with minimal processing complexity.

These developments highlight that RF lenses inherently map angular information into spatial power distributions, motivating the exploitation of such spatial focusing in RARE-based sensing; a direction that has not been previously explored. By incorporating an RF lens front-end, RARE-based sensing measurements can be converted into discriminative, AoA-dependent power signatures suitable for robust multi-user angular inference. Building on this insight, this paper introduces the multi-user Quantum-Power pROfile Based Estimation (PROBE) framework, naming motivated by~\cite{suk}, a lens-assisted RARE-based AoA estimation paradigm that operates entirely in the power domain without explicit phase recovery. The proposed framework leverages the RF lens-induced, AoA-dependent non-uniform power profiles formed across vapor cells to reliably extract angular information under diverse operating conditions. To systematically harness this structure, we develop a unified modeling and algorithmic framework that bridges quantum sensing physics with array signal processing. Our main contributions are summarized as follows:
\begin{itemize}
\item We establish a lens-embedded RARE reception model in which the multi-user RARE observation obtained by beam-propagation method (BPM) is shown to converge, after polarization averaging and snapshot-wise power accumulation, to a nonnegative superposition of AoA-dependent power-profile atoms determined primarily by the RF lens focusing response. This analysis reveals that angular information is inherently encoded in the power domain of RARE measurements through the lens-induced spatial power-focusing mechanism.
\item Building on the lens-embedded RARE power-profile analysis, we analytically derive an AoA-dependent power-profile dictionary and rigorously formulate multi-user AoA estimation as a sparse nonnegative representation problem in the power domain. Leveraging this structure, we develop two complementary Quantum-PROBE algorithms that explicitly solve the resulting estimation problem. Specifically:
\begin{enumerate}
\item A principled NN-LASSO-based framework that estimates a sparse nonnegative angular representation via an accelerated proximal-gradient method, followed by cluster-based AoA decoding, enabling accurate and robust multi-user AoA estimation with controlled computational complexity.
\item A low-complexity successive interference cancellation (SIC)-based framework that sequentially identifies and removes dominant angular components through cosine-similarity matching and nonnegative projection, thereby providing an efficient approximate solver with a favorable accuracy-complexity tradeoff suitable for large-scale RARE arrays.
\end{enumerate}%Leveraging this structure, we develop two complementary Quantum-PROBE algorithms: (i) a principled non-negative least absolute shrinkage and selection operator (NN-LASSO)-based approach that robustly recovers sparse angular representations and enables accurate multi-user AoA estimation with controlled computational complexity, and (ii) a low-complexity successive interference cancellation (SIC)-based scheme that iteratively identifies dominant angular components via cosine-similarity matching of lens-induced power profiles, achieving a favorable accuracy–complexity tradeoff suitable for large-scale RARE arrays.
\item We show that the proposed Quantum-PROBE frameworks achieve computational complexity linear with the number of RARE elements, rendering them considerably more computationally efficient than existing methods. Extensive simulations demonstrate that the proposed Quantum-PROBE frameworks consistently outperform existing RARE-based and conventional RF-based benchmarks, validating that the joint synergy between RF lens power focusing and quantum-level RARE sensitivity enables highly efficient AoA sensing under low-power operation.
\end{itemize}
\emph{The key novelty of this paper lies not in the RF lens or RARE individually, but in revealing how RF lens-induced power focusing fundamentally enables reliable multi-user AoA estimation from magnitude-only atomic measurements by transforming angular information into structured, AoA-dependent power profiles.}
\begin{figure}[t]
  \begin{center}
    \includegraphics[width=0.95\columnwidth,keepaspectratio]{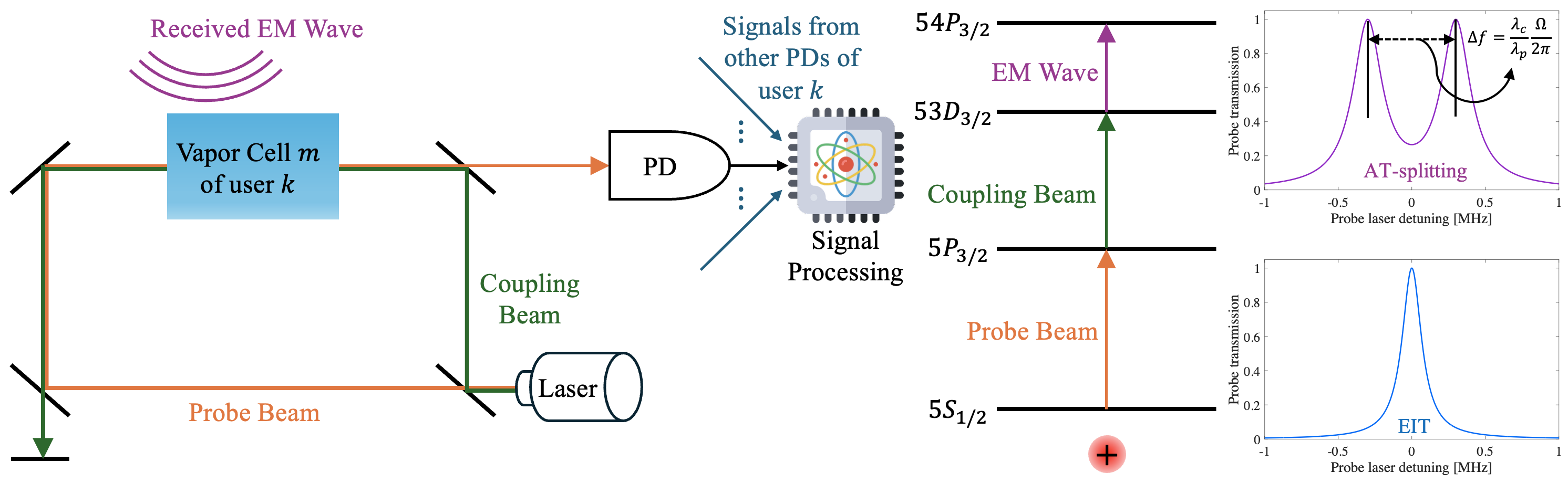}
    \caption{Illustration of the signal processing mechanism in an RARE. The impinging electromagnetic wave induces coupling between two highly excited Rydberg states (e.g., $53D_{3/2}$ and $54P_{3/2}$), leading to AT-splitting. The observed spectral separation $\Delta f$ is converted to the Rabi frequency $\Omega$ via~\eqref{rabidef}.}
    \label{fig_rabi}
  \end{center}
\end{figure}
\section{System Model}
\label{sec:quantum_probe}
\subsection{RARE-Based Reception Model}
\label{subsec:rar_model}
Consider an RARE comprising $M$ vapor cells, each containing an ensemble of Rydberg atoms. We consider $K_{\mathrm{U}}$ transmit sources (users) with unit-power symbol vector $\mathbf{s}=[s_1~\cdots~s_{K_{\mathrm{U}}}]^{\mathrm{T}}\in\mathbb{C}^{K_{\mathrm{U}}}$. The RF field incident on the $m$th cell is modeled as
\begin{equation}
  \mathbf{E}_m(t)= \sum_{k=1}^{K_{\mathrm{U}}}    \boldsymbol{\epsilon}_{m,k}     \sqrt{P_k}\rho_{m,k}s_k\cos(\omega t + \phi_{m,k}),
  \label{eq:xi_m}
\end{equation}
where $\boldsymbol{\epsilon}_{m,k}\in\mathbb{R}^3$ is the polarization vector at the $m$th cell corresponding to user $k$, $P_k$ is the transmit power of user $k$, and $\rho_{m,k}e^{j\phi_{m,k}}$ models the complex channel (path loss and phase) from user $k$ to cell $m$. We adopt a ULA-based RARE model with inter-cell spacing $d$, whereby the channel coefficient is expressed as
\begin{equation}
\label{eq:ula_chan}
\rho_{m,k}e^{j\phi_{m,k}}=\alpha_k e^{jk\frac{(m-1)d\sin\theta_k}{\lambda}},
\end{equation}
with $\theta_k$ and $\alpha_k\sim\mathcal{CN}(0,1)$ denoting the AoA and complex channel gain of user $k$, respectively~\cite{qesprit, qmusic, suk}. The standard ULA response is
\begin{equation}
\label{suar}
\mathbf{a}_{\mathrm{ULA}}(\theta)=[1~e^{j\frac{d\sin\theta}{\lambda}}~\cdots~e^{j\frac{(M-1)d\sin\theta}{\lambda}}]^{\mathrm{T}}\in\mathbb{C}^M.
\end{equation}

The incident RF field $\{\mathbf{E}_m\}$ couples with the Rydberg atomic transitions, giving rise to EIT and AT-splitting~\cite{efm}. The AT-splitting interval $\{\Delta f_m\}$ directly maps to the corresponding Rabi frequency $\{\Omega_m\}$~\cite{atomicjsac}:
\begin{equation}
  \Omega_m =  \left|  \sum_{k=1}^{K_{\mathrm{U}}}  \frac{1}{\hbar}  \boldsymbol{\mu}_{\mathrm{eg}}^{\mathrm{T}}\boldsymbol{\epsilon}_{m,k}
  \sqrt{P_k}\rho_{m,k} s_k e^{j\phi_{m,k}}  \right|~(m=1,\cdots,M),
  \label{eq:rabi_frequency_general}
\end{equation}
where $\boldsymbol{\mu}_{\mathrm{eg}}\in\mathbb{R}^3$ is the transition dipole moment vector. Moreover, by~\cite{efm},
\begin{equation}
\label{rabidef}
\Delta f_m=\frac{\lambda_c}{\lambda_p}\frac{\Omega_m}{2\pi}~(\forall m),
\end{equation}
where $\lambda_c$ and $\lambda_p$ denote the wavelengths of the coupling and probe beams, respectively. Hence, by sensing $\Delta f_m$, one can retrieve $\Omega_m$ via~\eqref{rabidef} and subsequently recover $\{s_k\}$ through~\eqref{eq:rabi_frequency_general}~\cite{qmusic, atomicjsac}. The overall signal processing procedure at RARE is depicted in Fig.~\ref{fig_rabi}.
\begin{figure}[t]
  \begin{center}
    \includegraphics[width=0.95\columnwidth,keepaspectratio]{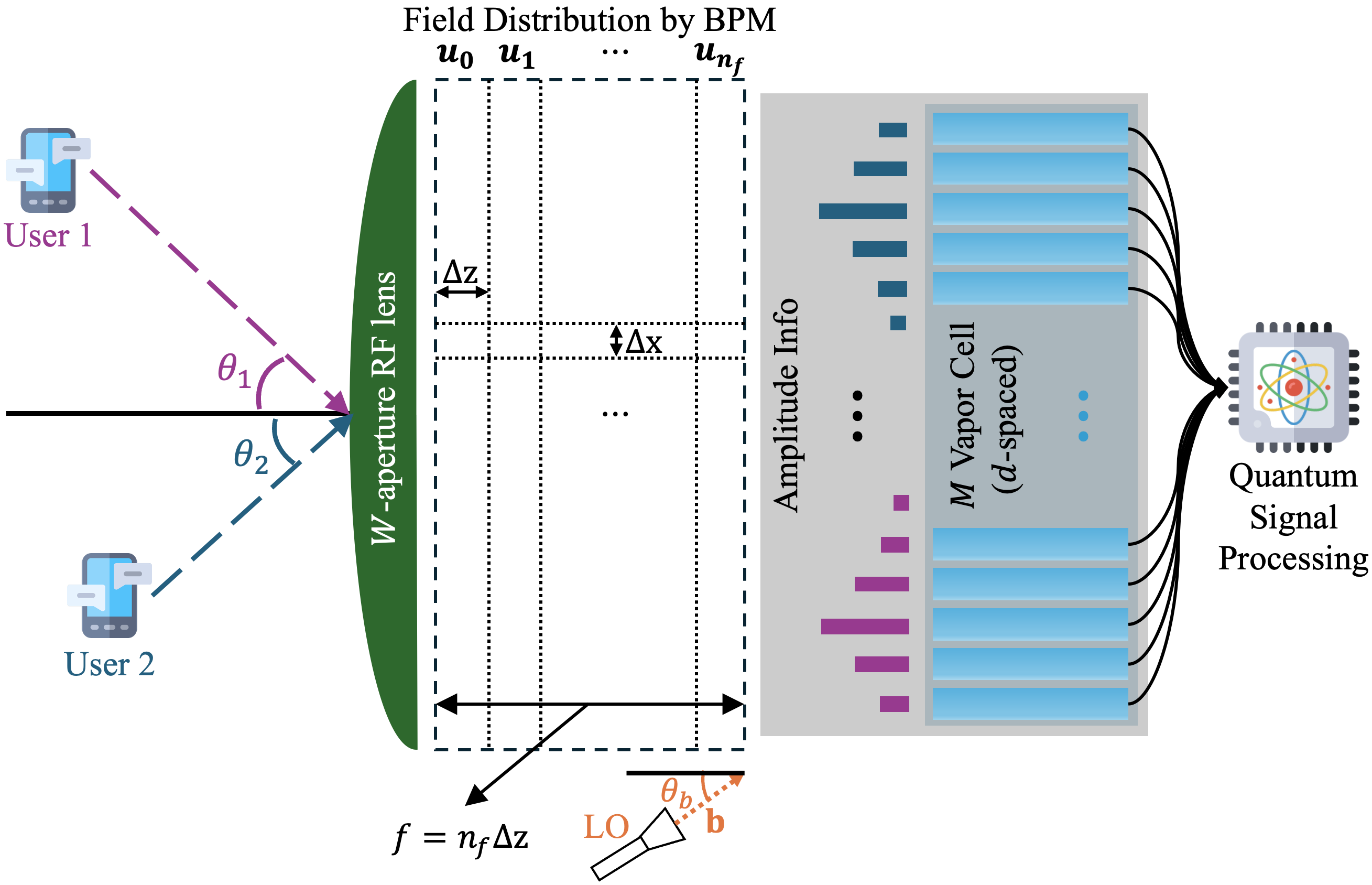}
    \caption{Illustration of the RARE-aided wireless system with an RF lens front-end.}
    \label{fig_sys}
  \end{center}
\end{figure}

By collecting the Rabi frequencies over $M$ cells, we define the magnitude-only RARE output vector
\begin{equation}
  \mathbf{y}  \triangleq [\Omega_1~\cdots~\Omega_M]^{\mathrm{T}}
  \in \mathbb{R}_+^M.
  \label{eq:rar_output_vector}
\end{equation}
To connect~\eqref{eq:rabi_frequency_general} with a classical MIMO notation, let
\begin{equation}
  a_{m,k}
  \triangleq
  \frac{1}{\hbar}
  \boldsymbol{\mu}_{\mathrm{eg}}^{\mathrm{T}}\boldsymbol{\epsilon}_{m,k}
  \sqrt{P_k}\rho_{m,k}e^{j\phi_{m,k}},
  \label{eq:channel_hmk}
\end{equation}
then~\eqref{eq:rabi_frequency_general} becomes
\begin{equation}
  \Omega_m
  =
  \left|\sum_{k=1}^{K_{\mathrm{U}}} a_{m,k} s_k\right|~ (m=1,\cdots,M).
  \label{eq:rabi_frequency_mimo}
\end{equation}
We denote the local-oscillator (LO) signal by $\mathbf b\in\mathbb{C}^M$ for phase recovery~\cite{atomicjsac}, whose $m$th component is
\begin{equation}
\label{bmcom}
b_m=\frac{1}{\hbar}\boldsymbol{\mu}_{\mathrm{eg}}^{\mathrm{T}}\boldsymbol{\epsilon}_{m,b} s_b\sqrt{P_b}\rho_{m,b}e^{j\phi_{m,b}},
\end{equation}
where $P_b,s_b,\rho_{m,b},\phi_{m,b}$, and $\boldsymbol{\epsilon}_{m,b}$ denote the known LO transmit power, unit-power LO symbol, path loss, phase shift, and polarization vector, respectively. Herein, we assume that the LO-to-RARE distance is sufficiently small so that the path loss is identical across cells~\cite{Precoding_atomicMIMO}, i.e., $\forall\rho_{m,b}=\beta$. Moreover, we consider the QSN $\mathbf{n}_q\sim\mathcal{CN}(\mathbf{0},\sigma_q^2\mathbf{I}_M)$~\cite{quanmobi, qsn, atomicjsac}. Defining the atomic MIMO channel $\mathbf{A}^*\in\mathbb{C}^{M\times K_{\mathrm{U}}}$ by $[\mathbf{A}^*]_{m,k}=a_{m,k}$, we obtain the compact form
\begin{equation}
  \mathbf{y} =  \bigl|\mathbf{A}^*\mathbf{s} + \mathbf{b}+\mathbf{n}_q\bigr|  \in\mathbb{R}_+^M,
  \label{eq:rar_magnitude_model}
\end{equation}
where the absolute value is applied elementwise. Equation~\eqref{eq:rar_magnitude_model} reflects that the RARE inherently discards phase information and encodes the field strength into the Rabi frequency magnitude, from which phase can be recovered by the biased GS algorithm~\cite{atomicjsac} or its modified version~\cite{qmusic}.

\begin{figure}[t]
  \centering
  \subfloat[]{%
    \includegraphics[width=0.24\textwidth]{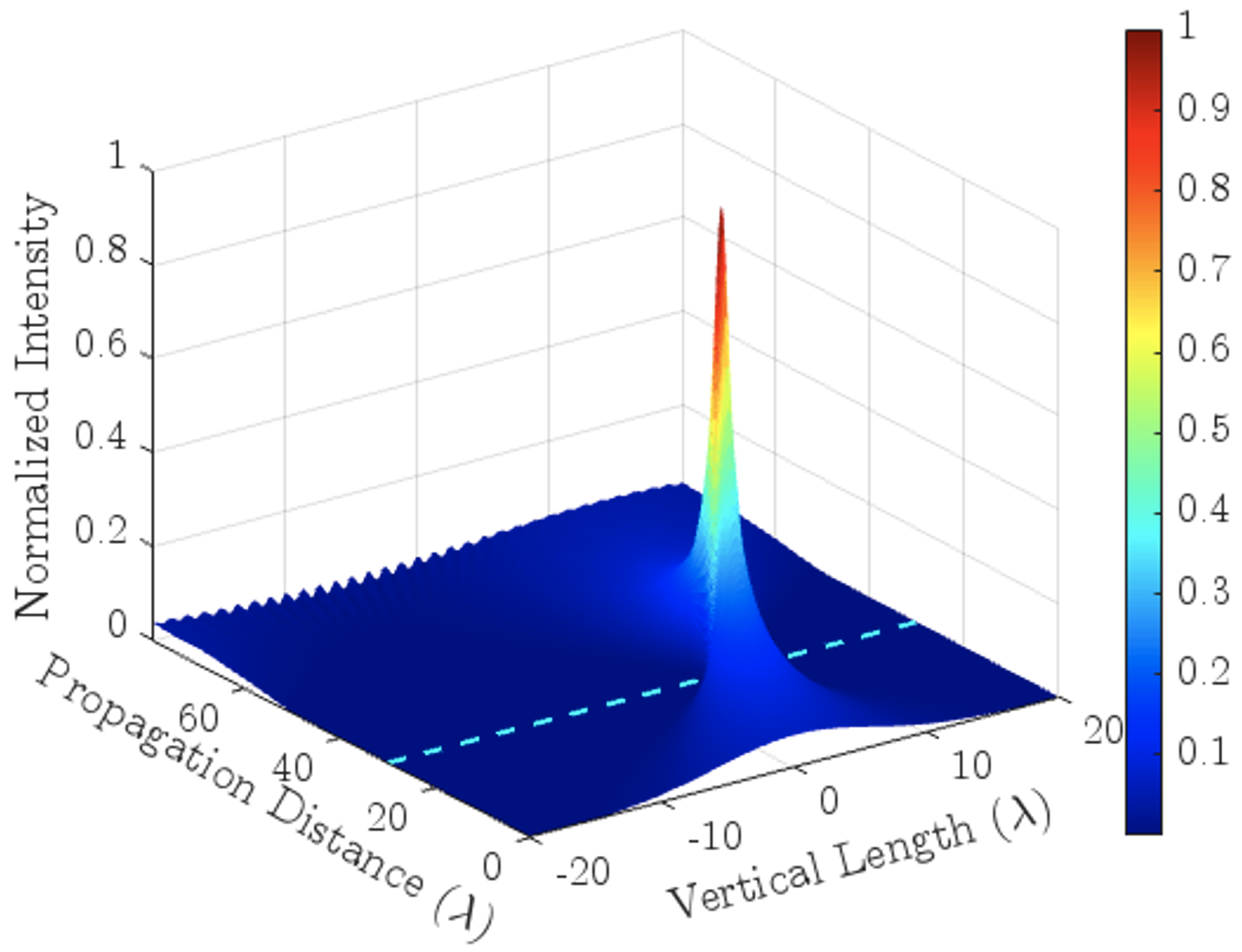}\label{fig_bpm12}%
  }
  \subfloat[]{%
    \includegraphics[width=0.24\textwidth]{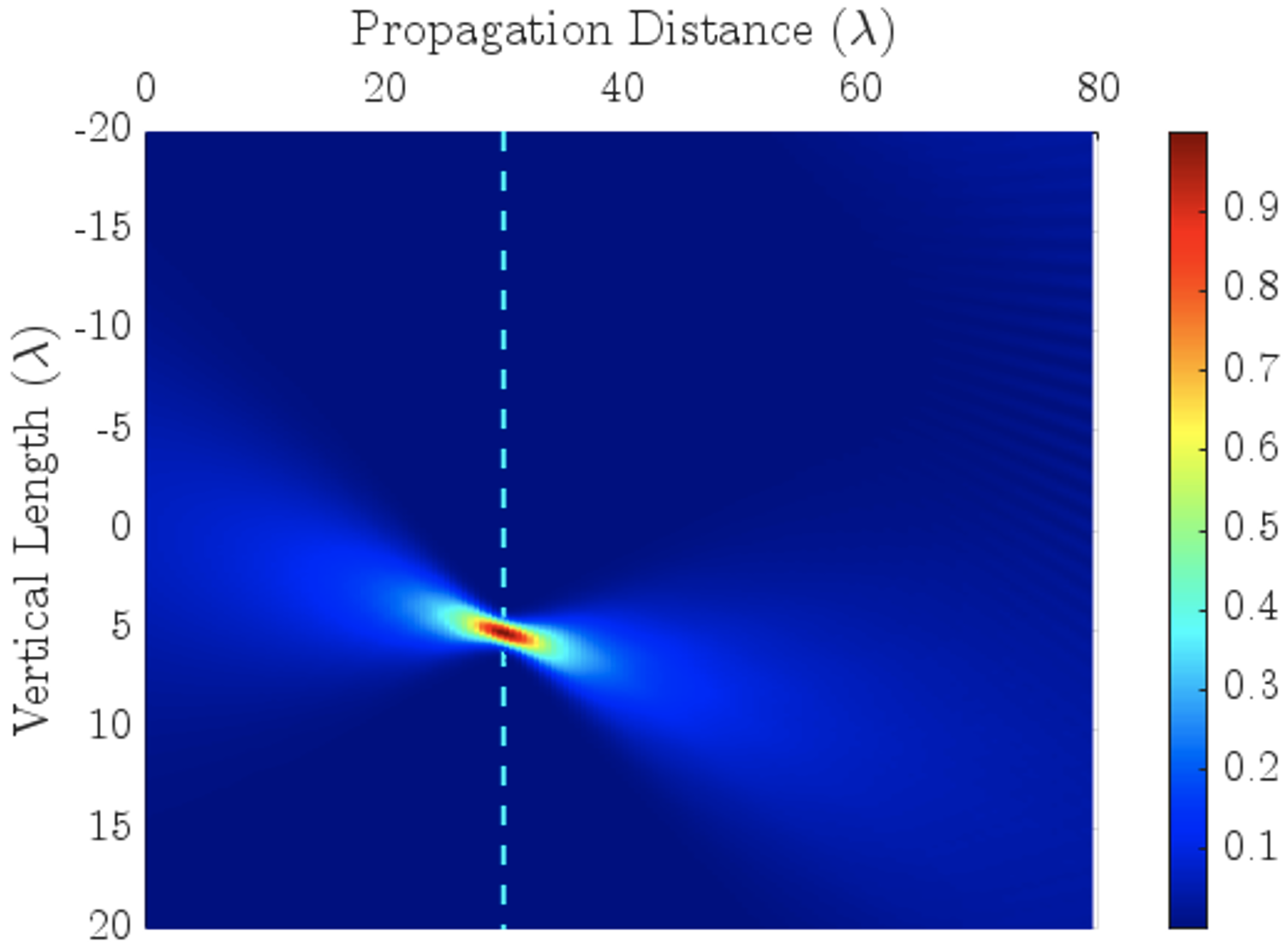}\label{fig_bpm22}%
  }
  \vfill
  \subfloat[]{%
    \includegraphics[width=0.24\textwidth]{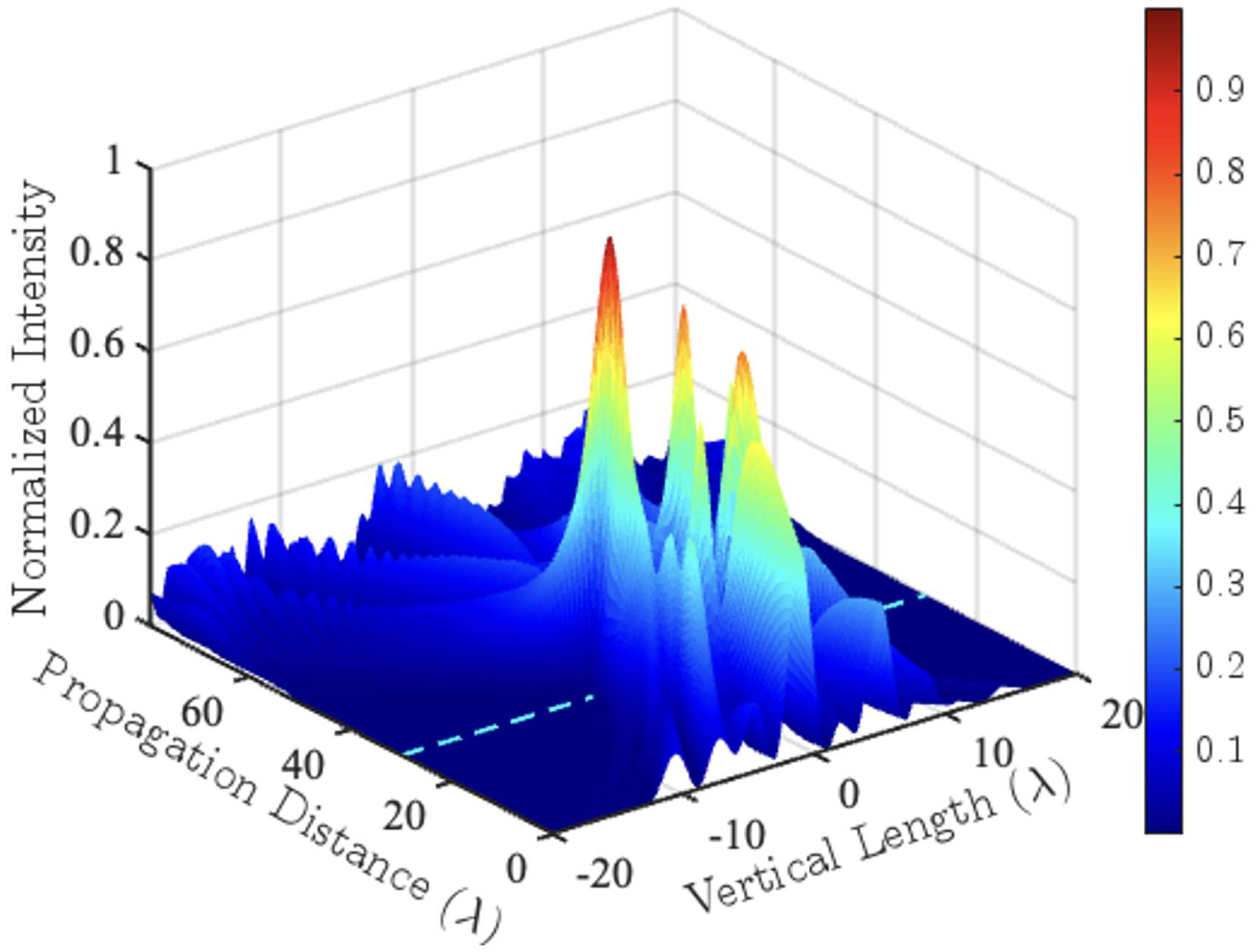}\label{fig_bpm1}%
  }
  \subfloat[]{%
    \includegraphics[width=0.24\textwidth]{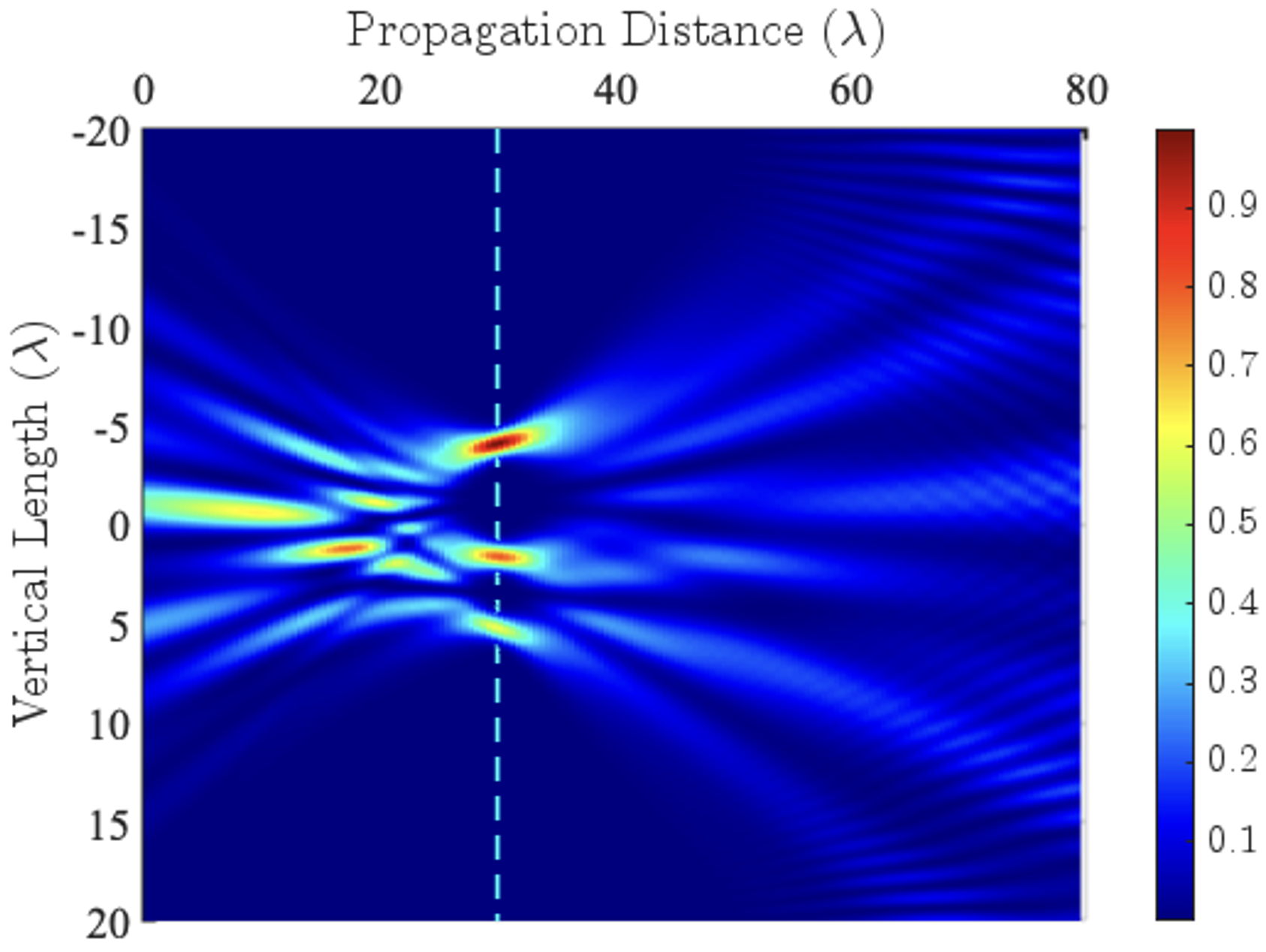}\label{fig_bpm2}%
  } 
  \caption{Normalized field distribution of the incident electromagnetic wave with (a)-(b) $\theta=10^\circ$ and (c)-(d) $\theta=-8^\circ, 3^\circ$, and $10^\circ$ at the RF lens, computed via BPM. The dashed line indicates $f$ from the lens along the propagation axis, where the RARE array is deployed.}
  \label{fig_bpm}
\end{figure}
\section{Analysis of Lens-Embedded Array Response}
\label{sec:framework}
\subsection{Lens-Embedded Array Response with BPM}
\label{subsec:lens_bpm}
As illustrated in Fig.~\ref{fig_sys}, we consider an RARE equipped with an RF lens front-end~\cite{lensntn, lensmeta}. The lens, characterized by an aperture width $W$ and focal length $f$, is positioned directly in front of a ULA-structured RARE. Herein, $f$ is given by $f^{-1}=\frac{n-1}{R}$~\cite{goodman2005introduction}, where $n=\sqrt{\varepsilon_r}$ denotes the refractive index determined by the dielectric relative permittivity $\varepsilon_r$, and $R$ represents the radius of curvature of the lens surface. The lens aperture is sampled with spacing $\Delta x$, yielding $N_s=\frac{W}{\Delta x}$ discrete spatial samples along the lens-horizontal axis. Let $\mathbf{u}_n\in\mathbb{C}^{N_s}$ denote the discrete field distribution at distance $z=n\Delta z$ from the lens, where $\Delta z$ is the propagation step along the lens-vertical axis. In the Fresnel diffraction regime, the BPM yields~\cite{kwon2016rf_LE, suk, goodman2005introduction}
\begin{equation}
\begin{aligned}
  \mathbf{u}_n
  &=
  \frac{e^{j k \Delta z}}{j\lambda \Delta z}
  \mathcal{F}^{-1}
  \Bigl\{
    \mathcal{F}\{\mathbf{u}_{n-1}\} \circ \mathbf{h}_{\mathrm{sys}}
  \Bigr\},\\
  h_{\mathrm{sys}}(p)
  &\triangleq
  \mathcal{F}
  \left\{
    e^{\frac{jk \bigl((p-1-\frac{N_s-1}{2})\Delta x\bigr)^2}{2\Delta z}}
  \right\},\\
  u_0(p)
  &\triangleq
  e^{-jk\frac{
      \bigl((p-1-\frac{N_s-1}{2})\Delta x\bigr)^2
      +2f(p-1-\frac{N_s-1}{2})\sin\theta    }{2f}},
\end{aligned}
\label{eq:bpm_recursion}
\end{equation}
where $k=\frac{2\pi}{\lambda}$ is the wavenumber, $\circ$ denotes the elementwise product, and $\mathcal{F}\{\cdot\}$ and $\mathcal{F}^{-1}\{\cdot\}$ are the Discrete Fourier Transform (DFT) and inverse DFT, respectively. The BPM field distribution for single and multi-user scenario is illustrated in Fig.~\ref{fig_bpm}.

Let $n_f$ be the index such that $z=n_f\Delta z\approx f$. Denote the focal-plane field distribution by $\mathbf u_f\triangleq \mathbf u_{n_f}\in\mathbb{C}^{N_s}$. The $m$th vapor cell lies at
\begin{equation}
  x_m=\Bigl(m-1-\frac{M-1}{2}\Bigr)d~ (m=1,\cdots,M),
  \label{eq:vcpc}
\end{equation}
and the discrete aperture coordinates are
\begin{equation}
  x_p=\Bigl(p-1-\frac{N_s-1}{2}\Bigr)\Delta x~ (p=1,\cdots,N_s).
  \label{lensdc}
\end{equation}
Mapping the vapor-cell positions to BPM indices yields
\begin{equation}
  p(m)  =  \operatorname{round}\left(\frac{x_m}{\Delta x}+\frac{N_s+1}{2}\right)~ (m=1,\cdots,M),
  \label{pmpm}
\end{equation}
where $p(m)\in\{1,\cdots,N_s\}$ is ensured by design. Hence, the lens-embedded array response for AoA $\theta$ is obtained by sampling $\mathbf u_f$ at $\{p(m)\}$:
\begin{equation}
  \mathbf{a}_{\mathrm{lens}}(\theta)
  =
  [u_f(p(1))~\cdots~u_f(p(M))]^{\mathrm{T}}
  \in\mathbb{C}^M.
  \label{eq:alens_def}
\end{equation}
In contrast to the conventional ULA response~\eqref{suar} (unit magnitudes and linear phase progression), $\mathbf{a}_{\mathrm{lens}}(\theta)$ exhibits a highly non-uniform magnitude pattern due to RF lens power focusing, as shown in Fig.~\ref{fig_bpm12} and~\ref{fig_bpm1}. This non-uniform, AoA-dependent signature gives rise to a new AoA inference framework in the RARE domain.
\subsection{Lens-Embedded RARE Power Profile: Multi-User Extension}
\label{subsec:rar_lens_profile_multiuser}
We now extend the lens-embedded RARE power-profile construction to the multi-user uplink. Let the true AoAs be $\boldsymbol\theta\triangleq[\theta_1~\cdots~\theta_{K_{\rm U}}]^{\mathrm T}$. Using~\eqref{eq:channel_hmk} together with the lens response~\eqref{eq:alens_def}, the lens-aided atomic channel coefficient from user $k$ to cell $m$ is
\begin{equation}
\label{eq:amk_multi}
a_{m,k}(\boldsymbol\theta)=\frac{1}{\hbar}\boldsymbol\mu_{\rm eg}^{\mathrm T}\boldsymbol\epsilon_{m,k}\sqrt{P_k}\alpha_k u_f\bigl(p(m);\theta_k\bigr).
\end{equation}
Define the complex baseband superposition at cell $m$ as
\begin{equation}
\label{eq:xm_multi}
x_m(\boldsymbol\theta)
\triangleq
\sum_{k=1}^{K_{\rm U}} a_{m,k}(\boldsymbol\theta)s_k,~\mathbf x(\boldsymbol\theta)\triangleq
[x_1(\boldsymbol\theta)~\cdots~x_M(\boldsymbol\theta)]^{\mathrm T}\in\mathbb{C}^{M}.
\end{equation}
In the absence of LO and noise, the Rabi magnitude at cell $m$ is by same means with~\eqref{eq:rabi_frequency_mimo}:
\begin{equation}
\label{eq:Omega_m_bar_multi}
\bar{\Omega}_m(\boldsymbol\theta)\triangleq |x_m(\boldsymbol\theta)|,~\bar{\boldsymbol\Omega}(\boldsymbol\theta)
\triangleq
|\mathbf x(\boldsymbol\theta)|
\in\mathbb{R}_+^M.
\end{equation}
Our framework exploits an AoA-dependent power profile across vapor cells. To this end, consider the elementwise squared magnitude
\begin{equation}
\label{esm}
\bar{\boldsymbol\Omega}(\boldsymbol\theta)^{\circ2}=|\mathbf x(\boldsymbol\theta)|^{\circ2}\in\mathbb{R}_+^{M}.
\end{equation}
\subsubsection{Conditional second moment and vanishing cross-terms}
\label{subsubsec:second_moment_multi}

In general,~\eqref{esm} contains cross-terms among users. To obtain a robust AoA-dependent signature, we consider the second moment of $x_m(\boldsymbol\theta)$. Assume that for each user $k$ and any cell $m$, $\{\boldsymbol\epsilon_{m,k}\}$ are independent across $m$ and $k$ and follow the isotropic polarization model uniform on the unit-circle (uniform angle on $[0,2\pi)$) orthogonal to $\hat{\mathbf k}(\theta_k)$ with $\|\hat{\mathbf k}(\theta_k)\|_2=1$, i.e., $\boldsymbol\epsilon_{m,k}^{\mathrm T}\hat{\mathbf k}(\theta_k)=0$ and $\|\boldsymbol\epsilon_{m,k}\|_2=1$~\cite{atomicjsac, Precoding_atomicMIMO}:
\begin{equation}
\label{eq:eet_multi}
\mathbb{E}\left[\boldsymbol\epsilon_{m,k}\boldsymbol\epsilon_{m,k}^{\mathrm T}\right]
=
\frac{1}{2}\Bigl(\mathbf I_3-\hat{\mathbf k}(\theta_k)\hat{\mathbf k}(\theta_k)^{\mathrm T}\Bigr),
\end{equation}
which implies the polarization-averaged projection gain
\begin{equation}
\label{eq:eta_k_def}
\eta(\theta_k)
\triangleq
\mathbb{E}\left[\big|\boldsymbol\mu_{\rm eg}^{\mathrm T}\boldsymbol\epsilon_{m,k}\big|^2\right]=\frac{1}{2}\left(\|\boldsymbol\mu_{\rm eg}\|_2^2-\big(\boldsymbol\mu_{\rm eg}^{\mathrm T}\hat{\mathbf k}(\theta_k)\big)^2\right).
\end{equation}
Then, expanding $|x_m|^2$ gives
\begin{equation}
\label{eq:expand_xm_sq}
\begin{aligned}
|x_m(\boldsymbol\theta)|^2
&=
\left|\sum_{k=1}^{K_{\rm U}} a_{m,k}(\boldsymbol\theta)s_k\right|^2\\
&=\sum_{k=1}^{K_{\rm U}} |a_{m,k}(\boldsymbol\theta)|^2+\sum_{\substack{k,\ell=1\\k\neq \ell}}^{K_{\rm U}}
a_{m,k}(\boldsymbol\theta)a_{m,\ell}^*(\boldsymbol\theta) s_k s_\ell^*.
\end{aligned}
\end{equation}
Taking $\mathbb E_{\alpha,\epsilon}[\cdot | \boldsymbol\theta,\mathbf s]$ yields
\begin{equation}
\begin{aligned}
&\mathbb E_{\alpha,\epsilon}\left[|x_m(\boldsymbol\theta)|^2| \boldsymbol\theta,\mathbf s\right]\\
&=\sum_{k=1}^{K_{\rm U}}  
\mathbb E_{\alpha,\epsilon} \left[|a_{m,k}(\boldsymbol\theta)|^2| \boldsymbol\theta\right]\\
&+
\sum_{\substack{k,\ell=1\\k\neq \ell}}^{K_{\rm U}}
s_k s_\ell^* 
\mathbb E_{\alpha,\epsilon} \left[a_{m,k}(\boldsymbol\theta)a_{m,\ell}^*(\boldsymbol\theta)| \boldsymbol\theta\right].
\label{eq:Ex2_split_det}
\end{aligned}
\end{equation}
For the cross term ($k\neq\ell$), substituting~\eqref{eq:amk_multi} gives
\begin{equation}
\begin{aligned}
&\mathbb E_{\alpha,\epsilon} \left[a_{m,k}(\boldsymbol\theta)a_{m,\ell}^*(\boldsymbol\theta)| \boldsymbol\theta\right]\\
&=
\frac{\sqrt{P_kP_\ell}}{\hbar^2} 
u_f \bigl(p(m);\theta_k\bigr) u_f^* \bigl(p(m);\theta_\ell\bigr) 
\\
&~~~~\mathbb E_{\alpha} \left[\alpha_k\alpha_\ell^*\right] \mathbb E_{\epsilon} \left[
(\boldsymbol\mu_{\rm eg}^{\mathrm T}\boldsymbol\epsilon_{m,k})
(\boldsymbol\mu_{\rm eg}^{\mathrm T}\boldsymbol\epsilon_{m,\ell})^*
\right],
\label{eq:cross_factorization_det}
\end{aligned}
\end{equation}
where we used the (inter-user) independence/uncorrelatedness between $\{\alpha_k\}$ and $\{\boldsymbol\epsilon_{m,k}\}$. Under either (i) inter-user uncorrelated fading $\mathbb E[\alpha_k\alpha_\ell^*]=0$ for $k\neq\ell$, or (ii) isotropic polarization with $\mathbb E[\boldsymbol\epsilon_{m,k}]=\mathbf 0$ and independence across $k$, the expectatiion-product in~\eqref{eq:cross_factorization_det} becomes zero, hence:
\begin{equation}
\mathbb E_{\alpha,\epsilon} \left[a_{m,k}(\boldsymbol\theta)a_{m,\ell}^*(\boldsymbol\theta)| \boldsymbol\theta\right]=0~(k\neq \ell).
\label{eq:cross_zero_det}
\end{equation}
Therefore,
\begin{equation}
\mathbb E_{\alpha,\epsilon} \left[|x_m(\boldsymbol\theta)|^2| \boldsymbol\theta,\mathbf s\right]=\sum_{k=1}^{K_{\rm U}}  \mathbb E_{\alpha,\epsilon} \left[|a_{m,k}(\boldsymbol\theta)|^2| \boldsymbol\theta\right].
\label{eq:Ex2_no_cross_det}
\end{equation}
Finally, since
\begin{equation}
|a_{m,k}(\boldsymbol\theta)|^2=\frac{P_k|\alpha_k|^2}{\hbar^2} \big|\boldsymbol\mu_{\rm eg}^{\mathrm T}\boldsymbol\epsilon_{m,k}\big|^2 \big|u_f(p(m);\theta_k)\big|^2,
\end{equation}
we obtain
\begin{equation}
\mathbb E_{\alpha,\epsilon} \left[|x_m(\boldsymbol\theta)|^2| \boldsymbol\theta,\mathbf s\right]=\sum_{k=1}^{K_{\rm U}}
\frac{P_k }{\hbar^2} 
\eta(\theta_k) 
g_m(\theta_k),
\label{eq:Ex2_multi_det_final}
\end{equation}
where $g_m(\theta_k)\triangleq |u_f(p(m);\theta_k)|^2$.
\subsubsection{Multi-user mean power profile}
\label{subsubsec:power_profile_multiuser}
Stacking~\eqref{eq:Ex2_multi_det_final} across $m$ yields the multi-user {mean power profile}:
\begin{equation}
\label{eq:power_profile_multi}
\mathbf p_{\rm MU}(\boldsymbol\theta)\triangleq \mathbb{E} \left[|\mathbf x(\boldsymbol\theta)|^{\circ2}| \boldsymbol\theta\right]=\sum_{k=1}^{K_{\rm U}} w_k \mathbf p_{\rm Q}(\theta_k),
\end{equation}
where by $\mathbf g(\theta)=[g_1 (\theta)\cdots g_M(\theta)]^{\mathrm{T}}$ and $w_k\triangleq \frac{P_k}{\hbar^2}$:
\begin{equation}
\label{eq:pQ_def_again}
\mathbf p_{\rm Q}(\theta)
\triangleq
\eta(\theta) |\mathbf a_{\rm lens}(\theta)|^{\circ2}
=
\eta(\theta) \mathbf g(\theta)\in\mathbb{R}_+^M.
\end{equation}
\begin{figure}[t]
  \centering
  \subfloat[]{%
    \includegraphics[width=0.24\textwidth]{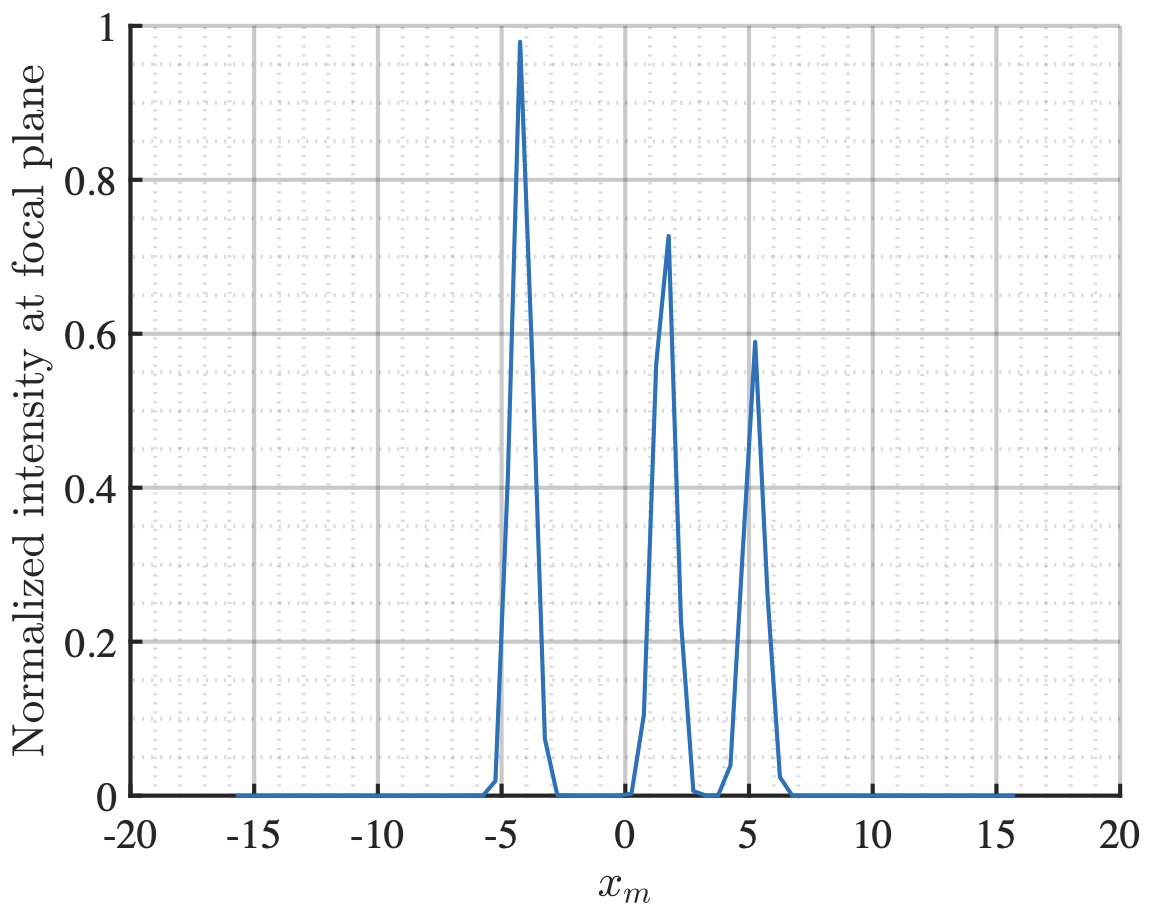}\label{fig_aoa1}%
  }
  \subfloat[]{%
    \includegraphics[width=0.24\textwidth]{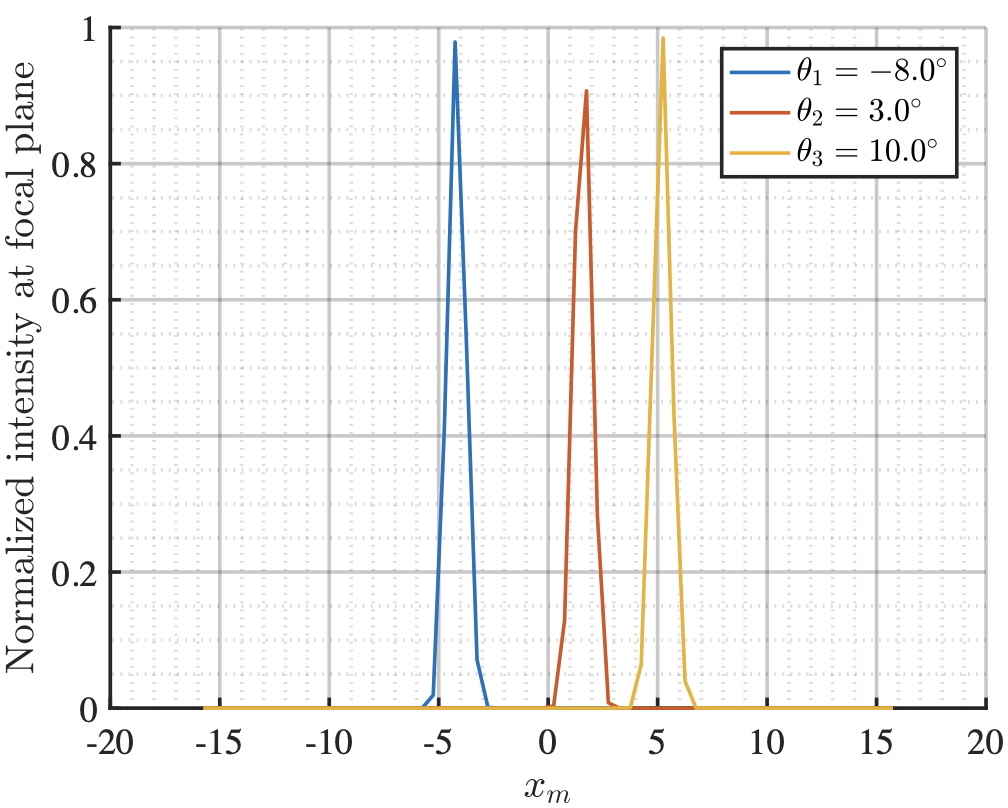}\label{fig_aoa2}%
  }
  \caption{Normalized (a) $\mathbf p_{\rm MU}(\boldsymbol\theta)$ for $\boldsymbol\theta=[-8^\circ, 3^\circ, 10^\circ]^{\mathrm{T}}$, and (b) the corresponding $\mathbf p_{\rm Q}(\theta_i)\in\mathcal{D}_{\mathrm{Q}}$ for $i=1,2,3$, normalized by the maximum value over the dictionary, which together constitute the profile shown in (a). The profiles in Fig.~\ref{fig_aoa1} are obtained by extracting the cross-section along the dashed line at the focal plane $f$ in Fig.~\ref{fig_bpm1}.}
  \label{fig_aoa}
\end{figure}
Equation~\eqref{eq:power_profile_multi} shows that, after averaging over symbol randomness and polarization, as shown in Fig.~\ref{fig_aoa1}, the multi-user RARE power profile with lens becomes a nonnegative superposition of $\{ \mathbf p_{\rm Q}(\theta_k)\}$ governed primarily by the RF lens BPM response. This perspective forms the basis of the proposed multi-user Quantum-PROBE algorithms, which recover the underlying AoAs by comparing the measured RARE power profile against the precomputed dictionary $\mathcal D_{\rm Q}$. Specifically, we discretize the AoA search region into a grid $\{\theta_i\}_{i=1}^{d_{\rm QP}}$ and precompute the corresponding lens-embedded RARE power profiles
\begin{equation}
\label{eq:dictionary_def_multi_rigorous}
\mathcal D_{\rm Q}
\triangleq
\left\{\mathbf p_{\rm Q}(\theta_i)=\eta(\theta_i)
\big|\mathbf a_{\rm lens}(\theta_i)\big|^{\circ 2}
\right\}_{i=1}^{d_{\rm QP}},
\end{equation}
which depend only on the lens geometry, array configuration, and atomic parameters, and can therefore be generated offline. Importantly, $\eta(\theta)$ only introduces a uniform scaling across vapor cells and does not alter the spatial shape of the lens-induced power profile. This property is exploited in the subsequent normalization and centering operations. Fig.~\ref{fig_aoa2} shows the corresponding dictionary element that form each power profile in Fig.~\ref{fig_aoa1}.
\subsubsection{Magnitude-only measurement with LO and QSN}
\label{subsubsec:measurement_multi}
At snapshot $k_{\rm snap}$, the magnitude-only RARE observation is
\begin{equation}
\label{eq:y_k_multi}
\mathbf y^{(k_{\rm snap})}=\left|\mathbf x^{(k_{\rm snap})}(\boldsymbol\theta)+\mathbf b^{(k_{\rm snap})}+\mathbf n_q^{(k_{\rm snap})}\right|\in\mathbb{R}_+^{M}.
\end{equation}
Define the time-averaged power profile over $K_{\rm snap}$ snapshots:
\begin{equation}
\label{eq:ybar_multi}
\bar{\mathbf y}
\triangleq
\frac{1}{K_{\rm snap}}
\sum_{k_{\rm snap}=1}^{K_{\rm snap}}
\left(\mathbf y^{(k_{\rm snap})}\right)^{\circ2}
\in\mathbb{R}_+^{M}.
\end{equation}
For each vapor cell $m$ and snapshot $k_{\rm snap}$, let $y_m^{(k_{\rm snap})}\triangleq |x_m^{(k_{\rm snap})}(\boldsymbol\theta)+b_m^{(k_{\rm snap})}+n_{q,m}^{(k_{\rm snap})}|.$ Then
\begin{equation}
\label{eq:zm_square_expand}
\begin{aligned}
&\bigl(y_m^{(k_{\rm snap})}\bigr)^2\\
&=
|x_m^{(k_{\rm snap})}|^2+|b_m^{(k_{\rm snap})}|^2+|n_{q,m}^{(k_{\rm snap})}|^2\\
&~~~~+2\Re\left\{x_m^{(k_{\rm snap})}\bigl(b_m^{(k_{\rm snap})}\bigr)^*\right\}
+2\Re\left\{x_m^{(k_{\rm snap})}\bigl(n_{q,m}^{(k_{\rm snap})}\bigr)^*\right\}\\
&~~~~+2\Re\left\{b_m^{(k_{\rm snap})}\bigl(n_{q,m}^{(k_{\rm snap})}\bigr)^*\right\}.
\end{aligned}
\end{equation}
Herein, since $\mathbf n_q^{(k_{\rm snap})}$ is independent of $(\mathbf x^{(k_{\rm snap})},\mathbf b^{(k_{\rm snap})})$ and satisfies $\mathbb{E}[\mathbf n_q^{(k_{\rm snap})}]=\mathbf 0$, we can get
\begin{equation}
\label{eq:noise_cross_vanish}
\mathbb{E}\left[x_m^{(k_{\rm snap})}\bigl(n_{q,m}^{(k_{\rm snap})}\bigr)^*| \boldsymbol\theta\right]
=
\mathbb{E}\left[b_m^{(k_{\rm snap})}\bigl(n_{q,m}^{(k_{\rm snap})}\bigr)^*\right]=0.
\end{equation}
Moreover, $\mathbf b^{(k_{\rm snap})}$ is statistically independent of $\mathbf x^{(k_{\rm snap})}(\boldsymbol\theta)$. Hence at snapshot $k_{\rm snap}$, write the user-induced field at cell $m$ as
\begin{equation}
\begin{aligned}
&x_m^{(k_{\rm snap})}(\boldsymbol\theta)
\\&=
\sum_{k=1}^{K_{\rm U}}
\underbrace{
\frac{1}{\hbar}
\boldsymbol\mu_{\rm eg}^{\mathrm T}\boldsymbol\epsilon_{m,k}^{(k_{\rm snap})}
\sqrt{P_k}\alpha_k^{(k_{\rm snap})}
u_f\bigl(p(m);\theta_k\bigr) s_k^{(k_{\rm snap})}
}_{\triangleq~x_{m,k}^{(k_{\rm snap})}},
\label{eq:xm_snap_def}
\end{aligned}
\end{equation}
and the LO component as
\begin{equation}
b_m^{(k_{\rm snap})}
=
\frac{1}{\hbar}
\boldsymbol\mu_{\rm eg}^{\mathrm T}\boldsymbol\epsilon_{m,b}^{(k_{\rm snap})}
\sqrt{P_b}\beta^{(k_{\rm snap})}  e^{j\varphi_b^{(k_{\rm snap})}}.
\label{eq:bm_snap_def}
\end{equation}
Then,
\begin{equation}
\begin{aligned}
&\mathbb{E}\left[x_m^{(k_{\rm snap})}(b_m^{(k_{\rm snap})})^*|\boldsymbol\theta\right]\\
&=
\sum_{k=1}^{K_{\rm U}}
\mathbb{E}\left[x_{m,k}^{(k_{\rm snap})}(b_m^{(k_{\rm snap})})^*|\boldsymbol\theta\right]
\\
&=
\sum_{k=1}^{K_{\rm U}}
\frac{\sqrt{P_kP_b}}{\hbar^2}
u_f\bigl(p(m);\theta_k\bigr)\mathbb{E}\Big[
\alpha_k^{(k_{\rm snap})}\big(\beta^{(k_{\rm snap})}\big)^*\\
&~~~~~~~~s_k^{(k_{\rm snap})}e^{-j\varphi_b^{(k_{\rm snap})}}\big(\boldsymbol\mu_{\rm eg}^{\mathrm T}\boldsymbol\epsilon_{m,k}^{(k_{\rm snap})}\big)
\big(\boldsymbol\mu_{\rm eg}^{\mathrm T}\boldsymbol\epsilon_{m,b}^{(k_{\rm snap})}\big)^*
 \Big| \boldsymbol\theta\Big].
\label{eq:xb_cross_expand}
\end{aligned}
\end{equation}
The LO-side random variables $\{\beta^{(k_{\rm snap})},\varphi_b^{(k_{\rm snap})},\boldsymbol\epsilon_{m,b}^{(k_{\rm snap})}\}$ are independent of the user-side random variables $\{\alpha_k^{(k_{\rm snap})},s_k^{(k_{\rm snap})},\boldsymbol\epsilon_{m,k}^{(k_{\rm snap})}\}$ for every $k$. Hence the expectation in~\eqref{eq:xb_cross_expand} factorizes as~\eqref{eq:factorize_xb}.
\begin{figure*}
\begin{equation}
\begin{aligned}
&\mathbb{E}\Big[\alpha_k^{(k_{\rm snap})}\big(\beta^{(k_{\rm snap})}\big)^*s_k^{(k_{\rm snap})}e^{-j\varphi_b^{(k_{\rm snap})}}\big(\boldsymbol\mu_{\rm eg}^{\mathrm T}\boldsymbol\epsilon_{m,k}^{(k_{\rm snap})}\big)\big(\boldsymbol\mu_{\rm eg}^{\mathrm T}\boldsymbol\epsilon_{m,b}^{(k_{\rm snap})}\big)^*\Big|\boldsymbol\theta\Big]
\\
&=
\mathbb{E}\left[\alpha_k^{(k_{\rm snap})} s_k^{(k_{\rm snap})}|\boldsymbol\theta\right]
\mathbb{E}\left[\big(\beta^{(k_{\rm snap})}\big)^*e^{-j\varphi_b^{(k_{\rm snap})}}\right]\mathbb{E}\left[\boldsymbol\mu_{\rm eg}^{\mathrm T}\boldsymbol\epsilon_{m,k}^{(k_{\rm snap})}\right]\mathbb{E}\left[\big(\boldsymbol\mu_{\rm eg}^{\mathrm T}\boldsymbol\epsilon_{m,b}^{(k_{\rm snap})}\big)^*\right].
\label{eq:factorize_xb}
\end{aligned}
\end{equation}
\hrule
\end{figure*}
Now, due to the isotopic property of $\boldsymbol\epsilon$: $\mathbb{E}\left[\boldsymbol\epsilon_{m,k}^{(k_{\rm snap})}\right]=\mathbb{E}\left[\boldsymbol\epsilon_{m,b}^{(k_{\rm snap})}\right]=\mathbf 0$, which implies
\begin{equation}
\mathbb{E}\left[\boldsymbol\mu_{\rm eg}^{\mathrm T}\boldsymbol\epsilon_{m,k}^{(k_{\rm snap})}\right]
=
\boldsymbol\mu_{\rm eg}^{\mathrm T}\mathbb{E}[\boldsymbol\epsilon_{m,k}^{(k_{\rm snap})}]
=0,~
\mathbb{E}\left[\big(\boldsymbol\mu_{\rm eg}^{\mathrm T}\boldsymbol\epsilon_{m,b}^{(k_{\rm snap})}\big)^*\right]=0.
\label{eq:mu_eps_zero}
\end{equation}
Substituting~\eqref{eq:mu_eps_zero} into~\eqref{eq:factorize_xb} yields
\begin{equation}
\mathbb{E}\left[x_m^{(k_{\rm snap})}(b_m^{(k_{\rm snap})})^*|\boldsymbol\theta\right]=0.
\label{eq:xb_cross_zero}
\end{equation}
Therefore, taking conditional expectation of~\eqref{eq:zm_square_expand} yields
\begin{equation}
\label{eq:second_moment_multi_rigorous}
\mathbb{E}\left[\bigl(y_m^{(k_{\rm snap})}\bigr)^2| \boldsymbol\theta\right]
=
\mathbb{E}\left[|x_m^{(k_{\rm snap})}(\boldsymbol\theta)|^2| \boldsymbol\theta\right]
+
\mathbb{E}\left[|b_m^{(k_{\rm snap})}|^2\right]
+\sigma_q^2.
\end{equation}
Now we are going to evaluate $\mathbb{E}[|b_m^{(k_{\rm snap})}|^2]$. By~\eqref{bmcom}:
\begin{equation}
\label{eq:bm_abs_square}
|b_m^{(k_{\rm snap})}|^2
=
\frac{P_b|\beta|^2}{\hbar^2}\bigl|\boldsymbol\mu_{\rm eg}^{\mathrm T}\boldsymbol\epsilon_{m,b}^{(k_{\rm snap})}\bigr|^2,
\end{equation}
and the LO polarization follows the same randomized model:
\begin{equation}
\label{eq:lo_pol_second_moment}
\mathbb{E}\left[\boldsymbol\epsilon_{m,b}^{(k_{\rm snap})}\bigl(\boldsymbol\epsilon_{m,b}^{(k_{\rm snap})}\bigr)^{\mathrm T}\right]
=
\frac{1}{2}\left(\mathbf I_3-\hat{\mathbf k}(\theta_b)\hat{\mathbf k}(\theta_b)^{\mathrm T}\right),
\end{equation}
where $\theta_b$ is the known LO AoA and $\hat{\mathbf k}(\theta_b)$ is the corresponding unit propagation direction.
Then by same means with~\eqref{eq:eta_k_def}:
\begin{equation}
\mathbb{E}\left[\bigl|\boldsymbol\mu_{\rm eg}^{\mathrm T}\boldsymbol\epsilon_{m,b}^{(k_{\rm snap})}\bigr|^2\right]
=\frac{1}{2}\left(\|\boldsymbol\mu_{\rm eg}\|_2^2-\bigl(\boldsymbol\mu_{\rm eg}^{\mathrm T}\hat{\mathbf k}(\theta_b)\bigr)^2\right)= \eta(\theta_b).
\label{eq:eta_theta_b_def}
\end{equation}
Substituting~\eqref{eq:lo_pol_second_moment}-\eqref{eq:eta_theta_b_def} into~\eqref{eq:bm_abs_square} gives the desired constant floor:
\begin{equation}
\label{eq:E_bm_square}
\mathbb{E}\left[|b_m^{(k_{\rm snap})}|^2\right]
=
\frac{P_b|\beta|^2}{\hbar^2}\eta(\theta_b)~(\forall m).
\end{equation}
Thus, by taking expectation in~\eqref{eq:zm_square_expand} and stacking with $m$:
\begin{equation}
\label{eq:E_y_square_vec}
\mathbb{E}\left[\bigl(\mathbf y^{(k_{\rm snap})}\bigr)^{\circ2}| \boldsymbol\theta\right]
=
\mathbf p_{\rm MU}(\boldsymbol\theta)
+
\frac{P_b|\beta|^2}{\hbar^2}\eta(\theta_b)\mathbf 1_M
+\sigma_q^2\mathbf 1_M,
\end{equation}
where the first term comes from~\eqref{eq:power_profile_multi}. Finally, by the strong law of large numbers applied to~\eqref{eq:ybar_multi},
\begin{equation}
\label{eq:measured_profile_multi_rigorous}
\begin{aligned}
&\bar{\mathbf y}\xrightarrow[]{K_{\rm snap}\to\infty}
\mathbf p_{\rm MU}(\boldsymbol\theta)
+\frac{P_b|\beta|^2}{\hbar^2}\eta(\theta_b)\mathbf 1_M
+\sigma_q^2\mathbf 1_M.
\end{aligned}
\end{equation}
\section{Proposed Multi-User Quantum-PROBE AoA Estimation}
\label{subsec:multiuser_qprobe}
\begin{figure*}[t]
  \begin{center}
    \includegraphics[width=1.98\columnwidth,keepaspectratio]{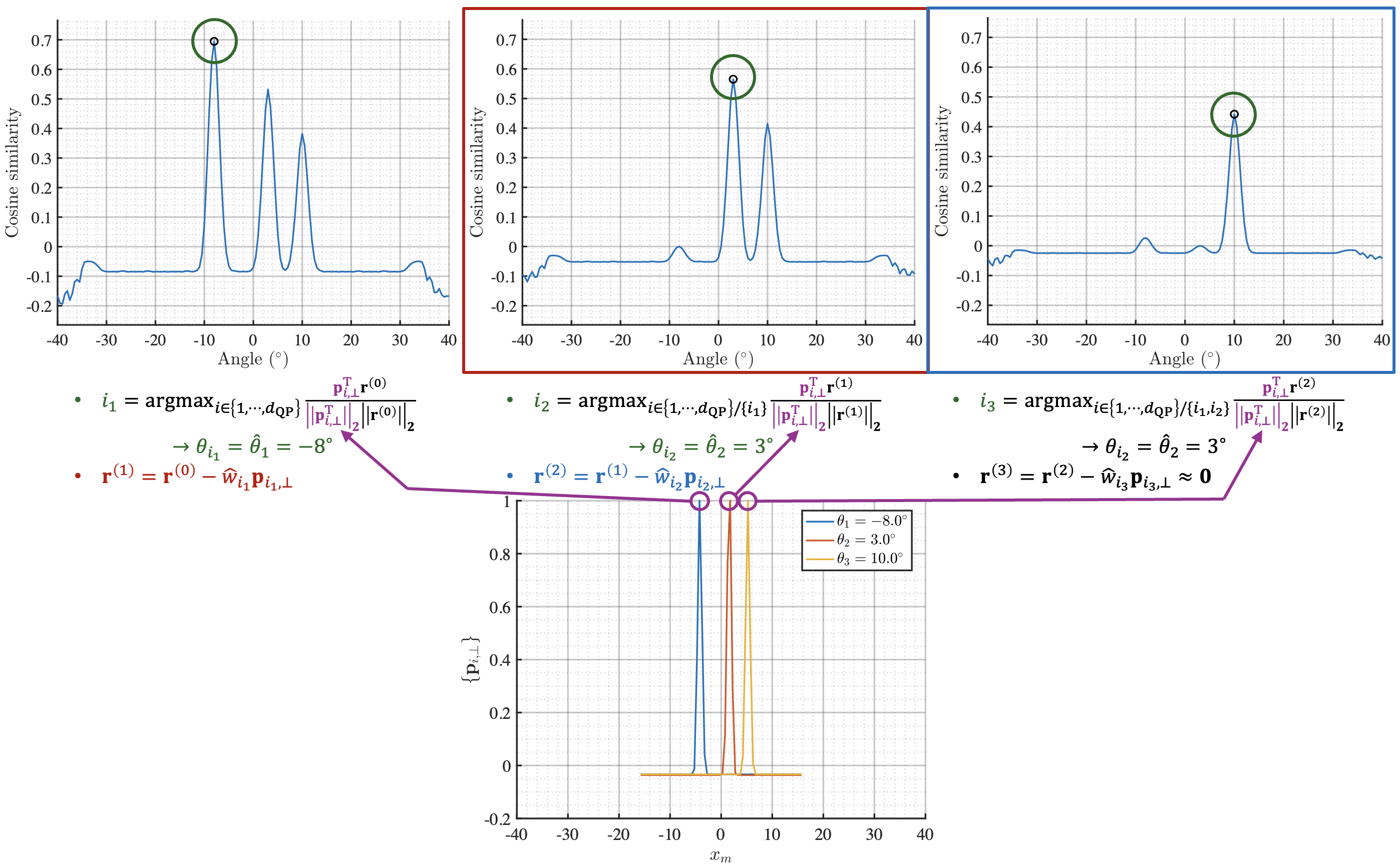}
    \caption{Illustration of Quantum-PROBE via SIC with $\boldsymbol\theta=[-8^\circ,3^\circ,10^\circ]^{\mathrm{T}}$.}
    \label{fig_sic}
  \end{center}
\end{figure*}

From Section~\ref{subsec:rar_lens_profile_multiuser}, the AoA estimation problem is therefore reduced to recovering the unknown $\{\theta_k\}_{k=1}^{K_{\rm U}}$ from a noisy observation of a {nonnegative superposition} of AoA-dependent $\{\mathbf p_{\rm Q}(\theta)\}$. We now formulate the framework of the proposed Quantum-PROBE algorithm. To start, since both the LO-induced floor and the QSN contribute only AoA-independent offsets proportional to $\mathbf 1_M$, we first apply the centering projector
\begin{equation}
\label{eq:Pi_def_multi_rigorous}
\mathbf\Pi
\triangleq
\mathbf I_M-\frac{1}{M}\mathbf 1_M\mathbf 1_M^{\mathrm T},~\mathbf\Pi\mathbf 1_M=\mathbf 0,
\end{equation}
to remove all constant components. Applying $\mathbf\Pi$ to~\eqref{eq:measured_profile_multi_rigorous} yields
\begin{equation}
\label{eq:y_perp_multi_rigorous}
\tilde{\mathbf y}_\perp
\triangleq\mathbf\Pi\bar{\mathbf y}\approx\sum_{k=1}^{K_{\rm U}} w_k\mathbf\Pi\mathbf p_{\rm Q}(\theta_k).
\end{equation}
Similarly, we define the centered dictionary:
\begin{equation}
\label{eq:atom_center_multi_rigorous}
\mathbf p_{i,\perp}
\triangleq
\mathbf\Pi\mathbf p_{\rm Q}(\theta_i),~\mathbf P_\perp
\triangleq
[\mathbf p_{1,\perp}~\cdots~\mathbf p_{d_{\rm QP},\perp}]\in\mathbb{R}^{M\times d_{\rm QP}},
\end{equation}
Notably,~\eqref{eq:atom_center_multi_rigorous} transforms $\mathcal{D}_{\mathrm{Q}}$ into the multiplied form ($\mathbf\Pi\mathbf p_{\rm Q}(\theta_k)$) in~\eqref{eq:atom_center_multi_rigorous} and removes all AoA-independent offsets, so that only the shape similarity induced by the RF lens focusing pattern is retained, as illustrated in the lower part of Fig.~\ref{fig_sic}.
\begin{figure}[t]
  \begin{center}
    \includegraphics[width=0.65\columnwidth,keepaspectratio]{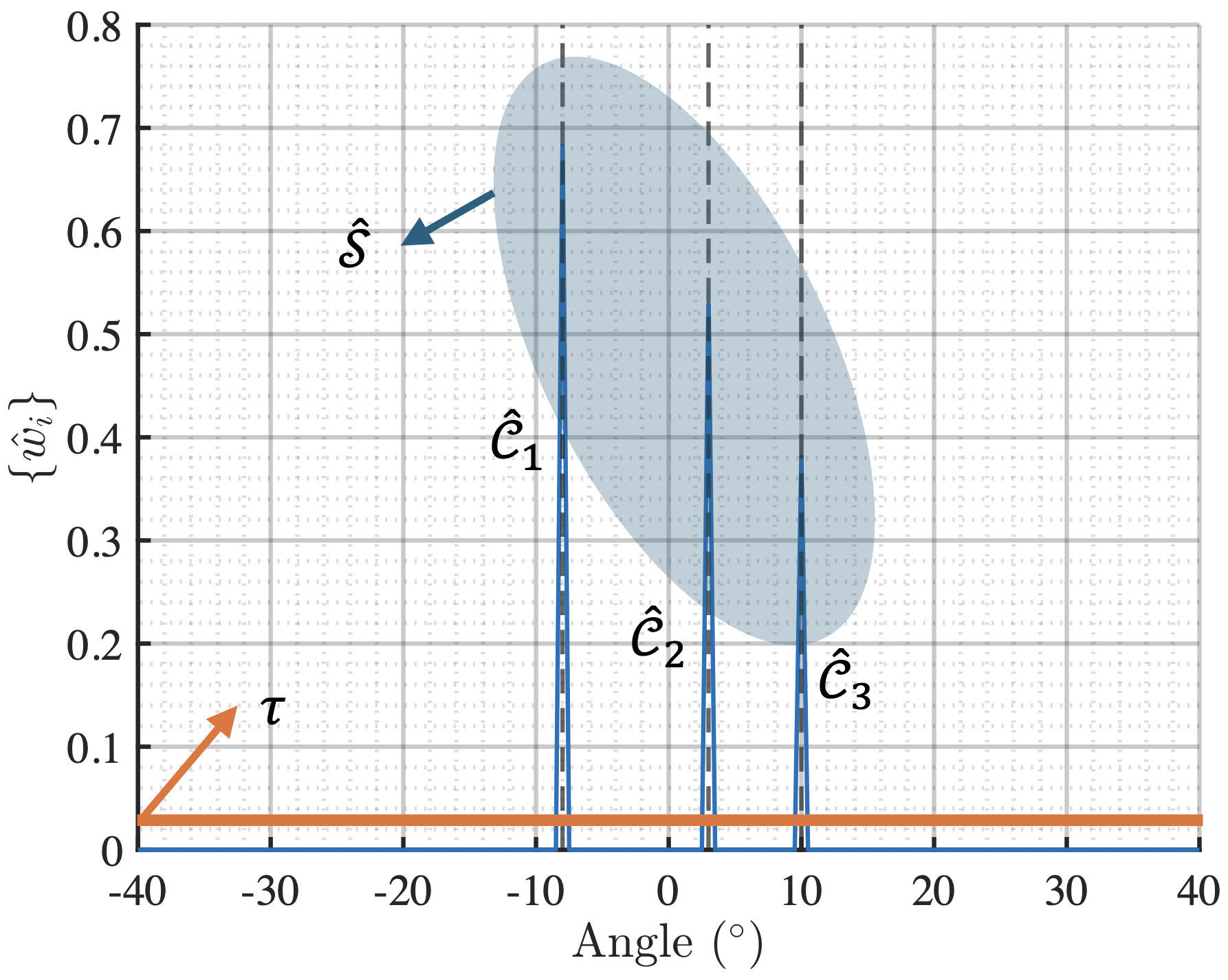}
    \caption{Illustration of $\{\hat w_i\}$ which generates $\hat{\mathcal S}$ and $\{\hat{\mathcal C}_c\}_{c=1}^{N_c}$ when $\boldsymbol\theta=[-8^\circ,3^\circ,10^\circ]^{\mathrm{T}}$.}
    \label{fig_hatw}
  \end{center}
\end{figure}
\subsection{Quantum-PROBE via NN-LASSO}
\label{subsubsec:nnls_multi_rigorous}
Combining~\eqref{eq:y_perp_multi_rigorous} and~\eqref{eq:atom_center_multi_rigorous},
the centered measurement admits the approximate linear model
\begin{equation}
\label{eq:lin_model_multi_rigorous}
\tilde{\mathbf y}_\perp\approx\mathbf P_\perp \bar{\mathbf w},~\bar{\mathbf w}\succeq \mathbf 0,
\end{equation}
where $\bar{\mathbf w}\triangleq[\bar{w}_1\cdots \bar{w}_{d_{\mathrm{QP}}}]^{\mathrm{T}}\in\mathbb{R}_+^{d_{\rm QP}}$ is a sparse nonnegative coefficient vector whose dominant entries are (approximately) $K_{\rm U}$ in number and whose support is located around the indices corresponding to $\{\theta_k\}_{k=1}^{K_{\rm U}}$. This representation reveals that multi-user AoA estimation in Quantum-PROBE can be interpreted as identifying a sparse nonnegative expansion of $\tilde{\mathbf y}_\perp$ over $\mathbf P_\perp$~\cite{nonne}, whose atoms encode AoA-dependent lens-induced power-focusing signatures.

Specifically, we estimate $\bar{\mathbf w}\in\mathbb R_+^{d_{\mathrm{QP}}}$ by solving the following NN-LASSO problem~\cite{nnlasso}:
\begin{equation}
\label{eq:nn_lasso}
\hat{\bar{\mathbf w}}
=\argmin_{\bar{\mathbf w}\succeq \mathbf 0}
\frac{1}{2}
\big\|
\tilde{\mathbf y}_\perp-\mathbf P_\perp \bar{\mathbf w}
\big\|_2^2+\lambda\|\bar{\mathbf w}\|_1,
\end{equation}
where $\lambda>0$ controls the sparsity level of the solution.

Problem~\eqref{eq:nn_lasso} is convex and admits an efficient first-order solution via a proximal-gradient method. Accordingly, we adopt the fast iterative shrinkage-thresholding algorithm (FISTA), which is an accelerated proximal-gradient method, to solve~\eqref{eq:nn_lasso}~\cite{fista}. Let $L=\|\mathbf P_\perp^{\mathrm T}\mathbf P_\perp\|_2$ denote the Lipschitz constant of the gradient of the quadratic term in~\eqref{eq:nn_lasso}. Starting from the initialization $\bar{\mathbf w}^{(0)}=\mathbf 0$, the update at iteration $t$ is given by~\cite{fista, fistaan}
\begin{equation}
\label{eq:fista_update}
\bar{\mathbf w}^{(t+1)}
=
\Big[
\mathcal S_{\lambda/L}
\Big(
\mathbf z^{(t)}
-
\frac{1}{L}
\mathbf P_\perp^{\mathrm T}
\big(
\mathbf P_\perp \mathbf z^{(t)}
-
\tilde{\mathbf y}_\perp
\big)
\Big)
\Big]_+,
\end{equation}
where $\mathcal S_{\tau}$ denotes the element-wise soft-thresholding operator:
\begin{equation}
\label{sto}
\mathcal S_{\tau}(x)
\triangleq
\mathrm{sign}(x)\max(|x|-\tau,0),
\end{equation}
and $[\cdot]_+=\max(\cdot,0)$ enforces the nonnegativity constraint. The auxiliary variable $\mathbf z^{(t)}$ is updated according to the standard FISTA acceleration rule. Specifically, letting $\eta^{(0)}=1$, after obtaining $\bar{\mathbf w}^{(t+1)}$ from~\eqref{eq:fista_update},
we update the acceleration parameter as
\begin{equation}
\label{eq:fista_eta}
\eta^{(t+1)}
\triangleq
\frac{1+\sqrt{1+4(\eta^{(t)})^2}}{2},
\end{equation}
and form the extrapolated point
\begin{equation}
\label{eq:fista_z}
\mathbf z^{(t+1)}
\triangleq
\bar{\mathbf w}^{(t+1)}
+
\frac{\eta^{(t)}-1}{\eta^{(t+1)}}
\big(\bar{\mathbf w}^{(t+1)}-\bar{\mathbf w}^{(t)}\big).
\end{equation}
Due to angular grid discretization and measurement noise, the nonzero entries of estimated $\hat{\mathbf w}$ typically appear as localized clusters around the true AoAs rather than isolated spikes. To robustly extract physical AoAs, we first identify the active support set
\begin{equation}
\label{eq:supp_rigorous}
\hat{\mathcal S}
\triangleq
\left\{
i :
\hat w_i > \tau
\right\},
\end{equation}
where $\tau>0$ is a small threshold introduced to suppress numerical artifacts. Next, $\hat{\mathcal S}$ is partitioned into $N_c$ connected components $\{\hat{\mathcal C}_c\}_{c=1}^{N_c}$ on the AoA grid, where two indices are considered connected if they correspond to adjacent angular bins. The generation of $\hat{\mathcal S}$ and $\{\hat{\mathcal C}_c\}_{c=1}^{N_c}$ is depicted in Fig.~\ref{fig_hatw}. Each $\hat{\mathcal C}_c$ is interpreted as the contribution of a single physical user, and the corresponding AoA $\bar{\theta}_c$ is estimated by a power-weighted centroid:
\begin{equation}
\label{eq:centroid_rigorous}
\bar{\theta}_c
\triangleq
\frac{
\sum_{i\in\hat{\mathcal C}_c}
\theta_i \hat w_i
}{
\sum_{i\in\hat{\mathcal C}_c}
\hat w_i
}~(c=1,\cdots,N_c).
\end{equation}
which provides a refined estimate that mitigates grid quantization error and leverages the smoothness of the lens-induced power peak. Finally, we rank the clusters according to their total masses $M_c\triangleq\sum_{i\in \hat{\mathcal C}_c}\hat w_i$ and select the $K_{\rm U}$ most significant ones as $\hat{\boldsymbol\theta}=[\hat{\theta}_1~\cdots~\hat{\theta}_{K_{\rm U}}]^{\mathrm T}.$:
\begin{equation}
\label{eq:topK_cluster_multi_rigorous}
\{\hat{\theta}_k\}_{k=1}^{K_{\mathrm U}}=
\operatorname*{Top\text{-}\it{K}_{\rm U}}_{c\in\{1, \cdots, N_c\}}
\left\{
\bar{\theta}_c; M_c
\right\}.
\end{equation}
\subsection{Quantum-PROBE via Low-Complexity SIC}
\label{subsubsec:sic_multi_rigorous}
To further reduce the computational burden of Quantum-PROBE by NN-LASSO, we propose a Quantum-PROBE by SIC scheme~\cite{isacjpark}. The key motivation is that $\{\mathbf p_{i,\perp}\}_{i=1}^{d_{\rm QP}}$ exhibits a strongly peaked and localized power-focusing pattern with respect to AoA, so that each active user contributes a dominant ``bump'' on the AoA grid. As a result, $\tilde{\mathbf y}_\perp$ can be well approximated by a sparse nonnegative superposition of a few atoms, shown in~\eqref{eq:power_profile_multi} and~\eqref{eq:measured_profile_multi_rigorous}, and the dominant component can be identified and removed iteratively.

To realize it, initialize the residual as $\mathbf r^{(0)}\triangleq \tilde{\mathbf y}_\perp$. For $\ell=1,\cdots,K_{\rm U}$, perform:
\begin{equation}
\begin{aligned}
&1): i_\ell
=
\argmax_{i\in\{1,\cdots,d_{\rm QP}\}/\{i_1, \cdots, i_{\ell-1}\}}\frac{ \mathbf p_{i,\perp}^{\mathrm T} \mathbf r^{(\ell-1)}}{\|\mathbf p_{i,\perp}\|_2 \|\mathbf r^{(\ell-1)}\|_2},\\
&2): \hat{\theta}_\ell\triangleq
\theta_{i_\ell},\\
&3): \hat w_{i_\ell}=\left[
\frac{\mathbf p_{i_\ell,\perp}^{\mathrm T}\mathbf r^{(\ell-1)}}
{\|\mathbf p_{i_\ell,\perp}\|_2^2}
\right]_+,\\
&4): \mathbf r^{(\ell)}=
\mathbf r^{(\ell-1)}-\hat w_{i_\ell} \mathbf p_{i_\ell,\perp}.
\label{eq:sic_update_multi_rigorous}
\end{aligned}
\end{equation}
Herein,~\eqref{eq:sic_update_multi_rigorous} consists of four procedures, each admitting a clear geometric interpretation.
\begin{enumerate}
\item{\textbf{Atom selection by maximum cosine similarity}:} The first line selects $i_\ell$ whose $\mathbf p_{i,\perp}$ is most aligned with $\mathbf r^{(\ell-1)}$ in the normalized inner-product sense. Since both $\mathbf p_{i,\perp}$ and $\mathbf r^{(\ell-1)}$ live in the centered subspace $\{\mathbf v:\mathbf 1_M^{\mathrm T}\mathbf v=0\}$, the cosine-similarity score $\frac{ \mathbf p_{i,\perp}^{\mathrm T} \mathbf r^{(\ell-1)}}{\|\mathbf p_{i,\perp}\|_2 \|\mathbf r^{(\ell-1)}\|_2}$ measures their ``shape similarity'' independent of scale, whose plot is depicted in the upper part of Fig.~\ref{fig_sic}. Excluding $\{i_1,\cdots,i_{\ell-1}\}$ prevents reselection of already-canceled components and encourages the algorithm to explain previously-unmodeled peaks.
\item{\textbf{AoA decoding:}} Given the selected $i_\ell$, the second line maps it back to the physical AoA grid by $\hat\theta_\ell\triangleq \theta_{i_\ell}$. This step is a direct dictionary lookup and simply translates the best-matching atom into the corresponding candidate AoA.
\item{\textbf{Nonnegative amplitude estimation via least-squares projection:}} Conditioned on $\mathbf p_{i_\ell,\perp}$, the third line estimates its contribution weight by a one-dimensional least-squares:
\begin{equation}
\label{odlf}
\min_{w\ge 0} \|\mathbf r^{(\ell-1)}-w\mathbf p_{i_\ell,\perp}\|_2^2,
\end{equation}
whose unconstrained minimizer is $\frac{\mathbf p_{i_\ell,\perp}^{\mathrm T}\mathbf r^{(\ell-1)}}{\|\mathbf p_{i_\ell,\perp}\|_2^2}$. Thereafter, $[\cdot]_+$ enforces the physics-consistent nonnegativity of the power-profile mixture coefficients, yielding $\hat w_{i_\ell}$ as the best nonnegative scalar explaining the residual along $\mathbf p_{i_\ell,\perp}$.
\item{\textbf{Residual update by orthogonal-component preservation:}} The fourth line subtracts the estimated component, $\mathbf r^{(\ell)}=\mathbf r^{(\ell-1)}-\hat w_{i_\ell}\mathbf p_{i_\ell,\perp}$, thereby removing the portion of the residual attributed to the $\ell$th detected user. Equivalently, when $\hat w_{i_\ell}$ is the (unconstrained) LS coefficient, this update is the standard projection of $\mathbf r^{(\ell-1)}$ onto the subspace orthogonal to $\mathbf p_{i_\ell,\perp}$. This is well justified in the lens-embedded setting, since each $\mathbf p_{i,\perp}$ exhibits a highly localized, AoA-dependent focusing pattern whose energy is concentrated around a small set of vapor cells and rapidly decays away from its peak. As a result, profiles associated with sufficiently separated AoAs have weak overlap (i.e., small normalized inner products), so subtracting $\hat w_{i_\ell} \mathbf p_{i_\ell,\perp}$ predominantly removes the contribution of the $\ell$th user while leaving the remaining users' peaks largely intact in the residual.
\end{enumerate}

Finally, the multi-user AoA estimate is returned as
\begin{equation}
\label{eq:sic_return_multi_rigorous}
\hat{\boldsymbol\theta}
=
[\hat{\theta}_1~\cdots~\hat{\theta}_{K_{\rm U}}]^{\mathrm T}.
\end{equation}
The overall procedure of Quantum-PROBE via SIC is illustrated in Fig.~\ref{fig_sic}, and the overall procedure of each framework is presented in Algorithm~\ref{alg:qp_nnlasso} and~\ref{alg:qp_sic_multi}; notably, both algorithms operate exclusively on lens-induced power profiles and do not require any phase information for AoA estimation.
\begin{algorithm}[t]
\caption{Quantum-PROBE via NN-LASSO}
\label{alg:qp_nnlasso}
\begin{algorithmic}[1]
\Require
$\{\mathbf y^{(k_{\rm snap})}\}_{k_{\rm snap}=1}^{K_{\rm snap}}, \mathbf P_\perp, L=\|\mathbf P_\perp^{\mathrm T}\mathbf P_\perp\|_2, K_{\rm U}, \lambda>0, \tau>0, \epsilon\ll1$
\State \textbf{Initialization:}
$\bar{\mathbf w}^{(0)}=\mathbf 0$, $\mathbf z^{(0)}=\mathbf 0$, $\eta^{(0)}=1$.
    \State Compute $\bar{\mathbf y}$ by~\eqref{eq:ybar_multi} and apply $\tilde{\mathbf y}_\perp \gets \mathbf\Pi \bar{\mathbf y}$.
\While {$\|\bar{\mathbf w}^{(t+1)}-\bar{\mathbf w}^{(t)}\|_2 > \epsilon$}
    \State Update $\mathbf{w}^{(t+1)}$ by~\eqref{eq:fista_update}.
    \State Update $\eta^{(t+1)}$ by~\eqref{eq:fista_eta}.
    \State Update $\mathbf{z}^{(t+1)}$ by~\eqref{eq:fista_z}.
    \State $t\leftarrow t+1$
\EndWhile
\State $\hat{\bar{\mathbf w}}\leftarrow \bar{\mathbf w}^{(t+1)}$
\State $\hat{\mathcal S}\leftarrow \{i:\hat w_i>\tau\}$
\State Partition $\hat{\mathcal S}$ into $\{\hat{\mathcal C}_c\}_{c=1}^{N_c}$.
\For{$c=1,2,\cdots,N_c$}
    \State $M_c \leftarrow \sum_{i\in\hat{\mathcal C}_c}\hat w_i$.
    \State Find $\hat\theta_c$ by~\eqref{eq:centroid_rigorous}.
\EndFor
\State Deduce $\hat{\boldsymbol\theta}$ by~\eqref{eq:topK_cluster_multi_rigorous}.
\State \Return $\hat{\boldsymbol\theta}$
\end{algorithmic}
\end{algorithm}

\begin{algorithm}[t]
  \caption{Quantum-PROBE via SIC}
  \label{alg:qp_sic_multi}
  \begin{algorithmic}[1]
    \Require
      $\{\mathbf y^{(k_{\rm snap})}\}_{k_{\rm snap}=1}^{K_{\rm snap}}, \mathbf P_\perp, K_{\rm U}$.
    \State Compute $\bar{\mathbf y}$ by~\eqref{eq:ybar_multi} and apply $\tilde{\mathbf y}_\perp \gets \mathbf\Pi \bar{\mathbf y}$.
    \State $\mathbf r^{(0)} \gets \tilde{\mathbf y}_\perp$.
    \For{$\ell = 0$ to $K_{\rm U}-1$}
      \State Select $i_\ell$ and corresponding $\hat\theta_\ell =\theta_{i_\ell}$ via~\eqref{eq:sic_update_multi_rigorous}.
      \State Estimate $\hat w_{\ell}$ via~\eqref{eq:sic_update_multi_rigorous}.
      \State Update $\mathbf r^{(\ell+1)}$ via~\eqref{eq:sic_update_multi_rigorous}.
    \EndFor
    \State \Return $\hat{\boldsymbol\theta}$
  \end{algorithmic}
\end{algorithm}
\subsection{Computational Complexity}
\label{subsec:complexity}
We analyze the computational complexity of the proposed multi-user Quantum-PROBE frameworks. The overall complexity is decomposed into the following components: (i) construction of the time-averaged RARE power profile, (ii) preprocessing via floor removal and centering, and (iii) multi-user AoA recovery via NN-LASSO or SIC. All dictionary-related quantities are assumed to be precomputed offline.
\subsubsection{Time-Averaged Power Profile Construction}
From~\eqref{eq:ybar_multi}, forming the time-averaged power profile requires one magnitude-squaring and accumulation per vapor cell and snapshot. Hence, the computational complexity is $\mathcal O\left(K_{\rm snap} M\right)$
\subsubsection{Floor Removal and Centering}
Centering is implemented using $\mathbf\Pi\mathbf v=\mathbf v-\frac{1}{M}(\mathbf 1_M^{\mathrm T}\mathbf v)\mathbf 1_M$, which requires one inner product and one vector subtraction. Therefore, the centering operation incurs $\mathcal O(M)$ computational cost.
\subsubsection{Quantum-PROBE via NN-LASSO with FISTA}
At each FISTA iteration in~\eqref{eq:fista_update}, the dominant computations are the matrix-vector products $\mathbf P_\perp\mathbf z^{(t)}$ and $\mathbf P_\perp^{\mathrm T}(\cdot)$, each costing $\mathcal O(Md_{\rm QP})$. The soft-thresholding, nonnegativity projection, and extrapolation steps incur $\mathcal O(d_{\rm QP})$. Hence, the per-iteration complexity of FISTA is $\mathcal O(M d_{\rm QP})$, and with $T$ iterations, the total complexity of NN-LASSO becomes $\mathcal O\left(T M d_{\rm QP}\right)$. 

After convergence of NN-LASSO, the post-processing stage consists of (i) support detection, (ii) connected-component clustering on the AoA grid, and (iii) weighted centroid computation. {Support detection} identifies $\hat{\mathcal S}=\{i:\hat w_i>\tau\}$ by a single thresholding pass over $\hat{\mathbf w}\in\mathbb{R}^{d_{\rm QP}}$, which requires $\mathcal O(d_{\rm QP})$ operations. {Connected-component clustering} is performed on a one-dimensional ordered AoA grid, where adjacent indices are grouped into connected components by a single linear scan with complexity of $\mathcal O(|\hat{\mathcal S}|)$. Centroid computation requires one pass over each cluster to evaluate $\sum_{i\in\hat{\mathcal C}_c}\hat w_i$ and $\sum_{i\in\hat{\mathcal C}_c}\theta_i\hat w_i$, whose total cost over all clusters is $\mathcal O(|\hat{\mathcal S}|)$.

Combining the iterative NN-LASSO solver and the post-processing stage, the overall computational complexity is
\begin{equation}
\label{nntotal}
\mathcal O\left(K_{\mathrm{snap}}M+T M d_{\rm QP} \right),
\end{equation}
where the post-processing-related term is typically negligible since $|\hat{\mathcal S}|\ll d_{\rm QP}$.
\subsubsection{Quantum-PROBE via SIC}
In Quantum-PROBE via SIC, after 1) and 2), each iteration requires evaluating $\mathbf p_{i,\perp}^{\mathrm T}\mathbf r^{(\ell-1)}$ for all $i=1,\cdots,d_{\rm QP}$, where each inner product costs $\mathcal O(M)$. Thus, each SIC iteration incurs $\mathcal O(M d_{\rm QP})$ computational complexity. Since SIC is performed for exactly $K_{\rm U}$ iterations, the total SIC complexity is $\mathcal O\left(K_{\rm U} M d_{\rm QP}\right)$. Combining the SIC solver and the post-processing stage, the overall computational complexity is
\begin{equation}
\label{sictotal}
\mathcal O\left(K_{\mathrm{snap}}M+K_{\rm U} M d_{\rm QP}\right).
\end{equation}

\begin{table}[t]
\centering
\caption{System Parameters}
\label{tabsim}
\begin{tabular}{l c}
\toprule
\textbf{Parameter} & \textbf{Value} \\
\midrule
Number of users $K_{\mathrm{U}}$ (unless referred) & 3 \\
Number of elements $M$ (unless referred) & 64 \\
Carrier frequency $f_c$ & 5~GHz\\
SNR (unless referred) & 5~dB \\
RARE spacing $d$ & $\frac{\lambda}{2}$\\
Number of snapshots $K_{\mathrm{snap}}$ & 1024 \\
Lens aperture $W$ & $40\lambda$\\
Focal length $f$ & $52\lambda$ \\
BPM grid $(\Delta x, \Delta z)$ & $\left(\frac{\lambda}{8}, \lambda \right)$\\
AoA distribution & $\mathcal{U}\left[-\frac{\pi}{12}, \frac{\pi}{12}\right]$\\
QSN \& JNTN $(\sigma_q^2, \sigma_t^2)$ & (-191, -176)~dBm\\
\bottomrule
\end{tabular}
\end{table}

Notably, both approaches scale linearly with $M$ and $d_{\rm QP}$. In contrast, classical subspace-based approaches such as MUSIC and ESPRIT typically require $\mathcal{O}(M^3)$ complexity for eigendecomposition, with MUSIC further incurring $\mathcal{O}(M^2d_{\mathrm{QP}})$ for grid-based spectral scanning. Even the RARE-adapted MUSIC variant in~\cite{qmusic} involves additional phase-recovery and bias-correction steps with at least $\mathcal{O}(M^3+MK^3+d_{\mathrm{QP}} M^2)$. Hence, both proposed Quantum-PROBE frameworks eliminate all cubic- and quadratic-in-$M$ operations and enable AoA inference with substantially lower real-time computational cost. Among them, Quantum-PROBE via SIC further accentuates this low-complexity advantage due to the smaller multiplicative factors preceding $M$. This behavior is analyzed and numerically verified in Section~\ref{roro}.
\section{Simulation Results}
In this section, we evaluate the AoA estimation accuracy of the proposed Quantum-PROBE frameworks and compare it with representative benchmarks. Unless otherwise specified, all simulations use the parameter settings summarized in Table~\ref{tabsim} with received signal-to-noise-ratio (SNR):
\begin{equation}
\label{resnr}
\mathrm{SNR}\triangleq\frac{\mathbb{E}\left[\big\|\mathbf{A}^*\mathbf s\big\|_2^2\right]} {\mathbb{E}\left[\|\mathbf n_q\|_2^2\right]}.
\end{equation}
For the atomic configuration, we employ the Rydberg transitions between $52D_{5/2}$ and $53P_{3/2}$ to detect an RF carrier at $f_c=5$~GHz. Following~\cite{rydpar}, the corresponding transition dipole moment is computed as $\boldsymbol{\mu}_{\mathrm{eg}}=[0,1785.916qa_0,0]^{\mathrm{T}}$, where $a_0=5.292\times10^{-11}$~m denotes the Bohr radius and $q=1.602\times10^{-19}$~C is the elementary charge. $\boldsymbol{\epsilon}_{m,b}$ and $\boldsymbol{\epsilon}_{m,k}$ are then independently drawn as unit-norm vectors lying on the circles orthogonal to their respective incident directions, while ${\phi_{m,b}}$ are generated uniformly over $[0,2\pi)$. For each operating point, we perform $10^3$ Monte Carlo realizations. We compare the proposed method with (i) the PROBE scheme expanded to multi-user scenario by detecting multiple peaks of the similarity based on a conventional RF receiver equipped with an RF lens~\cite{suk}, and (ii) the Quantum-MUSIC method~\cite{qmusic}, a MUSIC-type approach tailored to lens-free RARE reception. Herein, to ensure a fair comparison of sensing performance between atomic and conventional RF receivers, the power spectral densities of Johnson-Nyquist thermal noise (JNTN), denoted by $\sigma_t^2$, and QSN are evaluated at room temperature and adopted to determine the corresponding noise power~\cite{qsn, atomicjsac}. The root-mean-square error (RMSE) is defined as
\begin{equation}
\label{rmsed}
\mathrm{RMSE}\triangleq\sqrt{\frac{1}{TK}\sum_{t=1}^{T}\sum_{k=1}^{K}\big|\hat{\theta}_{t,k}-\theta_{t,k}\big|^{2}},
\end{equation}
where $T$ denotes the number of Monte Carlo realizations, $\theta_{t,k}$ is the true AoA of the $k$th source in the $t$th trial, and $\hat{\theta}_{t,k}$ is the corresponding estimate.

\begin{figure}[t]
  \centering
  \subfloat[]{%
    \includegraphics[width=0.24\textwidth]{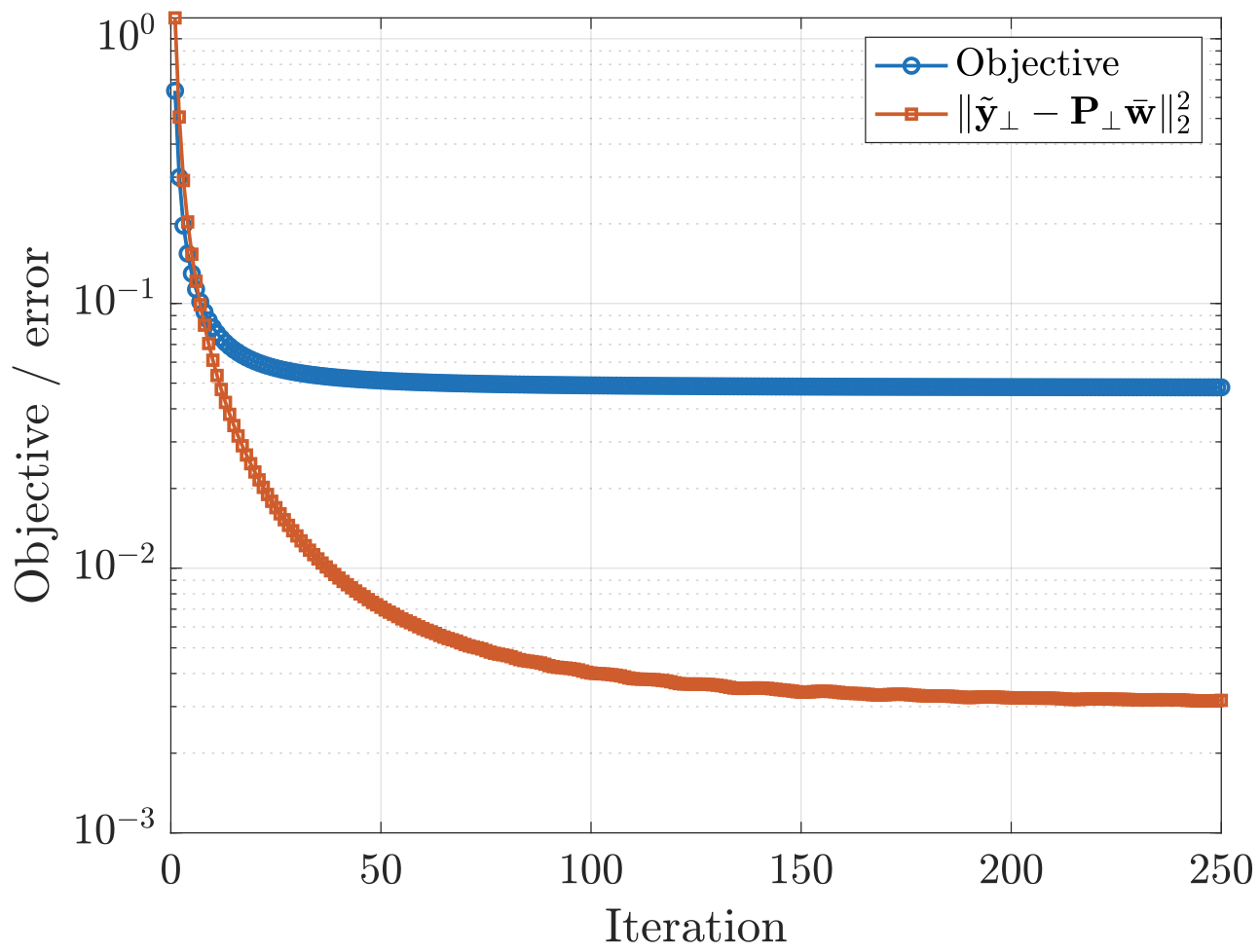}\label{fig_lassoc}%
  }
  \subfloat[]{%
    \includegraphics[width=0.24\textwidth]{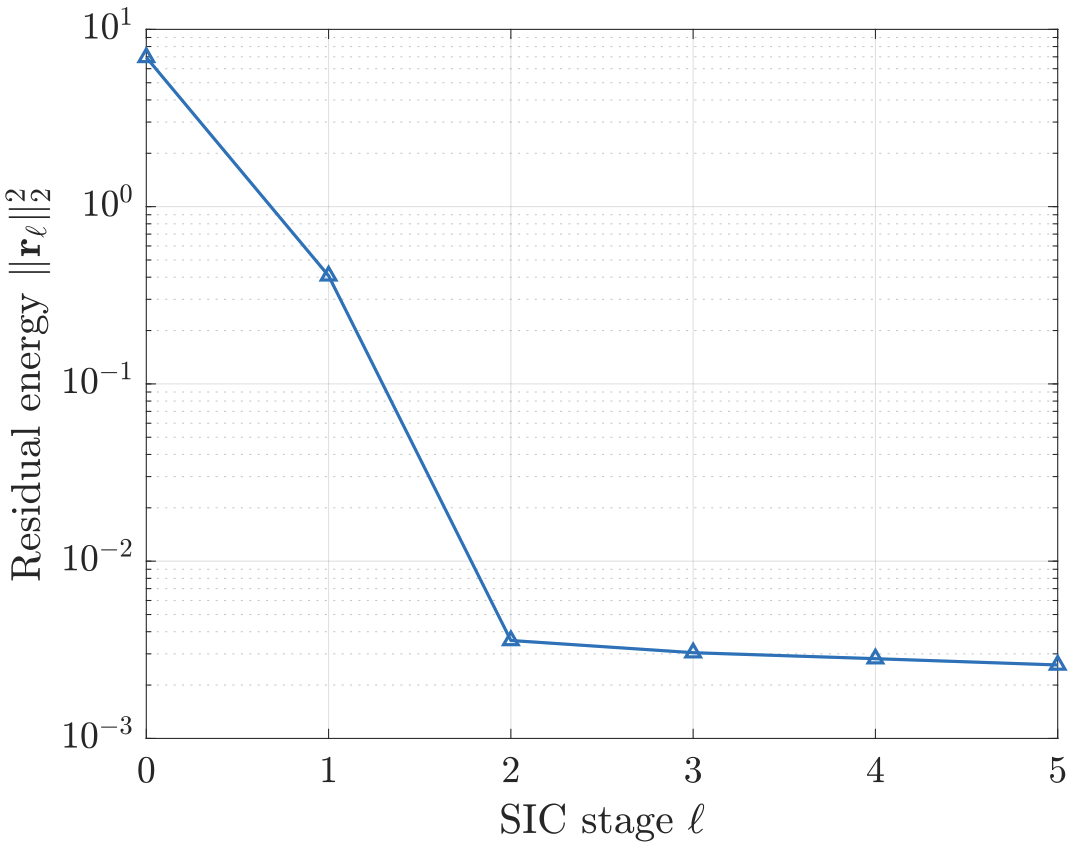}\label{fig_sicc}%
  }
  \caption{Convergence behavior of (a) the objective value in~\eqref{eq:nn_lasso} together with the modeling error $\|\tilde{\mathbf y}_\perp-\mathbf P_\perp \bar{\mathbf w}\|_2^2$ in~\eqref{eq:lin_model_multi_rigorous} over the iterations of Algorithm~\ref{alg:qp_nnlasso}, and (b) the residual energy $\|\mathbf r_\ell\|_2^2$ over $\ell$ in Algorithm~\ref{alg:qp_sic_multi}.}
  \label{fig_conv}
\end{figure}
\subsection{Reliability of Proposed Frameworks}\label{roro}
Fig.~\ref{fig_conv} confirms the favorable convergence behavior of the proposed algorithms. Specifically, Fig.~\ref{fig_lassoc} shows that both the objective value of~\eqref{eq:nn_lasso} and the modeling error $\|\tilde{\mathbf y}_\perp-\mathbf P_\perp \bar{\mathbf w}\|_2^2$ in~\eqref{eq:lin_model_multi_rigorous} decrease monotonically over the iterations of Algorithm~\ref{alg:qp_nnlasso} and rapidly stabilize after a moderate number of updates. This behavior is attributed to the convexity of the NN-LASSO subproblem together with the fixed, precomputed dictionary structure, which ensures that each iteration yields a non-increasing objective while progressively refining the approximation accuracy of the linearized model. Meanwhile, Fig.~\ref{fig_sicc} demonstrates that the residual energy $\|\mathbf r_\ell\|_2^2$ in Algorithm~\ref{alg:qp_sic_multi} decays sharply within the first few SIC stages and quickly reaches a low floor, indicating that the dominant AoA components are reliably identified and removed at early stages. The subsequent stages contribute only marginal refinements, reflecting the effective suppression of inter-user interference and error propagation. Overall, these results validate that the proposed NN-LASSO and SIC-based Quantum-PROBE algorithms exhibit stable and fast convergence, consistent with their underlying problem structures and update rules, and are well suited for practical implementation.

\begin{figure}[t]
  \begin{center}
    \includegraphics[width=0.65\columnwidth,keepaspectratio]{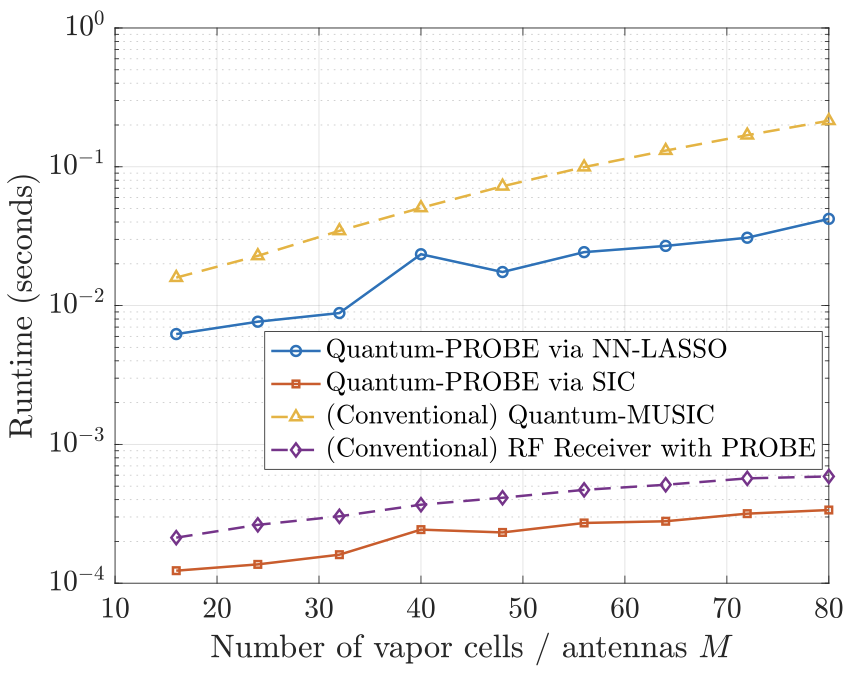}
    \caption{Algorithm runtime with respect to $M$, where for RF receivers, $M$ corresponds to the number of antennas.}
    \label{fig_compl}
  \end{center}
\end{figure}
Fig.~\ref{fig_compl} depicts the average runtime as a function of the number of vapor cells/antennas $M$ for different AoA estimation schemes. As $M$ increases, the proposed Quantum-PROBE methods exhibit a predictable and well-structured scaling behavior. In particular, the runtime of Quantum-PROBE via NN-LASSO remains higher than that of the SIC-based implementation. This is because NN-LASSO requires iterative optimization over the entire angular dictionary, including repeated gradient evaluations and projection steps to enforce nonnegativity and sparsity with $T$ iterations in~\eqref{nntotal}, which incur a non-negligible constant factor (around 100-200 in Fig.~\ref{fig_lassoc}) despite the favorable asymptotic scaling on $M$. By contrast, the SIC-based Quantum-PROBE achieves the lowest runtime and exhibits a much gentler increase with $M$, clearly highlighting its low-complexity nature. This favorable scaling stems from its linear dependence on $M$ and its reliance on only a small number of sequential residual updates, as illustrated in Fig.~\ref{fig_sicc}, together with matched-filtering operations for $K_{\mathrm{U}}$ users, as characterized in~\eqref{sictotal}. In comparison, Quantum-MUSIC suffers from substantially higher runtime that grows rapidly with $M$ due to the need for covariance matrix construction and eigenvalue decomposition, whose computational burden increases superlinearly ($\propto M^3$~\cite{qmusic}) with array size. Meanwhile, the conventional RF receiver with PROBE shows moderate runtime scaling but remains slightly slower than the SIC-based Quantum-PROBE.
\subsection{RMSE Performance Comparisons under Several Effects}
\begin{figure}[t]
  \begin{center}
    \includegraphics[width=0.9\columnwidth,keepaspectratio]{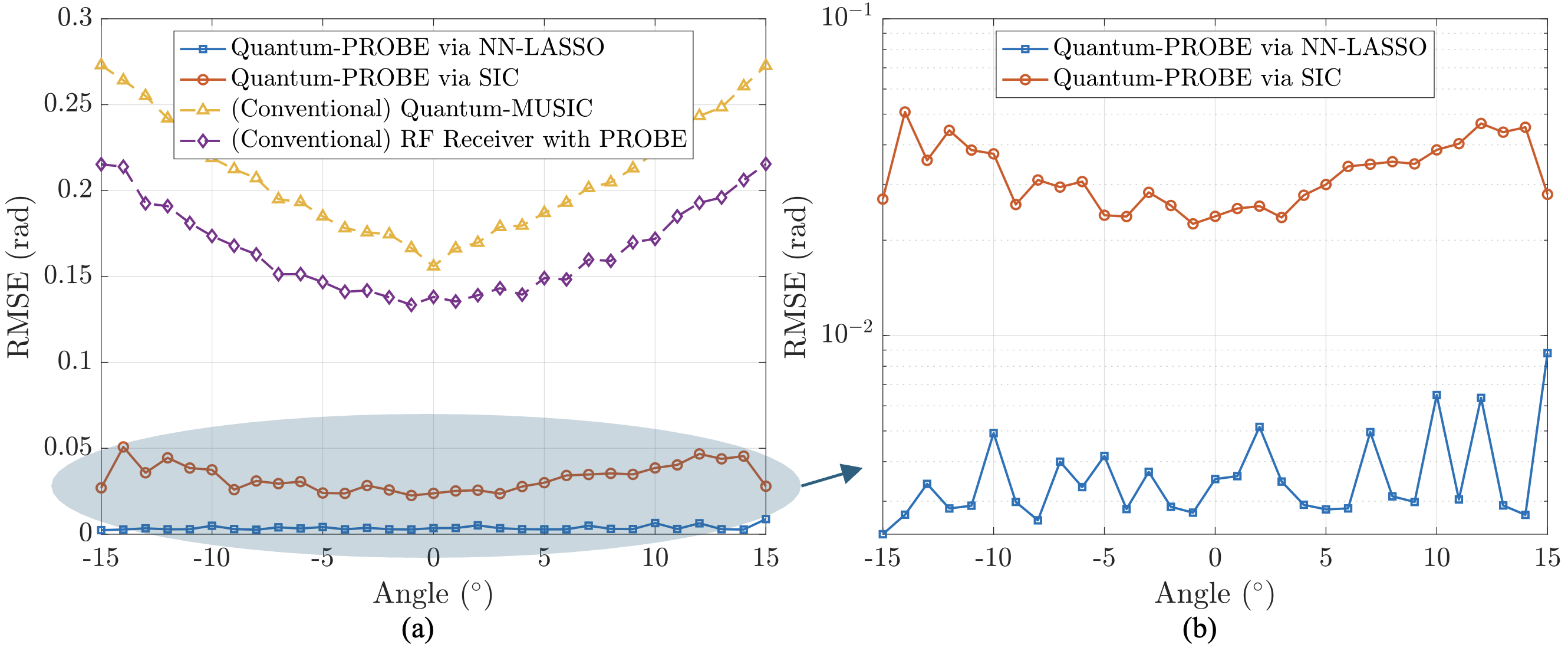}
    \caption{AoA RMSE with respect to varying AoAs.}
    \label{fig_ra}
  \end{center}
\end{figure}
Fig.~\ref{fig_ra} shows the RMSE versus the input AoA within the lens-supporting range. The estimation accuracy improves as the AoA approaches $0^\circ$ since the RF lens generates the most concentrated and symmetric power profile for near-broadside incidence~\cite{twcdoa, suk}. Herein, both proposed Quantum-PROBE via NN-LASSO and SIC frameworks maintain RMSE on the order of less than $10^{-2}$~rad near broadside, demonstrating robust and consistent performance compared to the benchmarks in the region where the lens focusing is most effective; the NN-LASSO variant achieves this accuracy at the expense of relatively higher computational complexity, as shown in Fig.~\ref{fig_compl}, while the SIC-based approach incurs a slight performance loss but offers a substantial complexity advantage; these performance trends among the benchmark schemes persist throughout the remaining simulations.

\begin{figure}[t]
  \begin{center}
    \includegraphics[width=0.65\columnwidth,keepaspectratio]{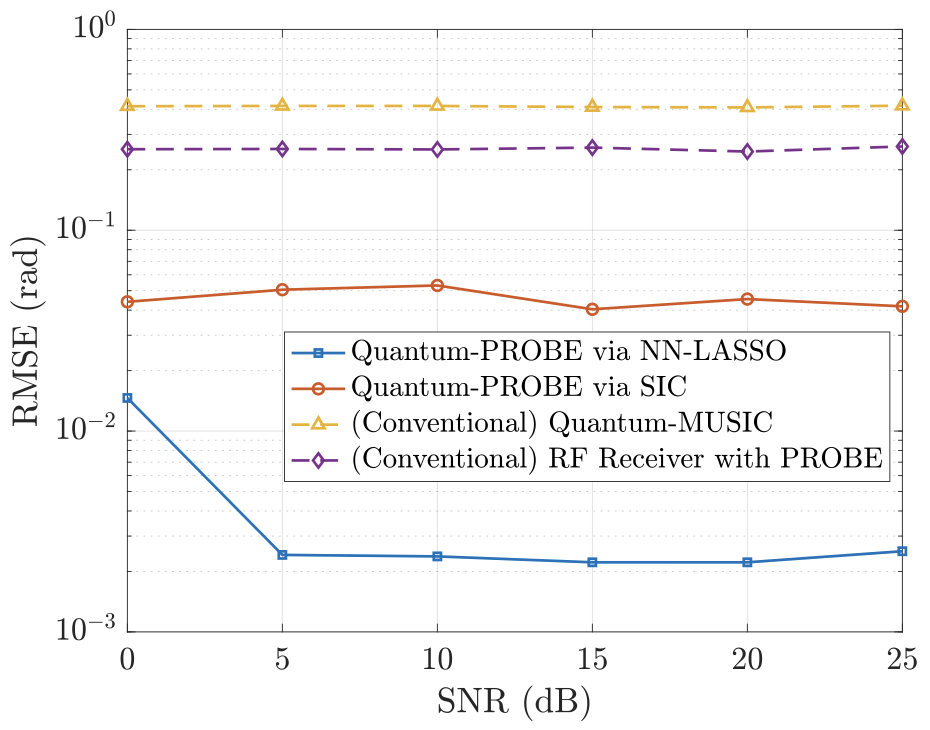}
    \caption{AoA RMSE with respect to SNR.}
    \label{fig_snr}
  \end{center}
\end{figure}
Fig.~\ref{fig_snr} depicts the RMSE performance as a function of SNR, clearly demonstrating that the proposed Quantum-PROBE frameworks consistently outperform all benchmark schemes under both recovery strategies. Specifically, at SNR=10~dB, Quantum-PROBE via SIC achieves an RMSE of approximately $5\times10^{-2}$~rad, which is about ten times lower than those attained by Quantum-MUSIC and the RF receiver with PROBE, while the Quantum-PROBE via NN-LASSO further reduces the RMSE to the order of $2\times10^{-3}$~rad, yielding more improvement over the baselines. This pronounced performance gain arises from the ability of the proposed framework to jointly exploit the intrinsic reception characteristics of RARE, including their lower QSN, together with the RF lens-induced, AoA-dependent spatial power-focusing effect. As a result, for a given SNR, the proposed frameworks inherently operate with lower absolute received power compared to classical RF receivers under the same SNR, yet still achieve superior AoA estimation accuracy compared to conventional RARE-based sensing when assisted by the RF lens, highlighting their enhanced sensing efficiency through joint synergy; these performance trends among the benchmark schemes persist throughout the remaining simulations.
%By contrast, Quantum-MUSIC relies on intermediate phase recovery using a biased GS algorithm followed by subspace estimation from magnitude-only observations, where residual phase ambiguities and reconstruction errors fundamentally limit the achievable angular resolution. Meanwhile, the RF receiver with PROBE is adversely affected by higher RF noise levels, which destabilize the lens-induced power patterns and degrade AoA discriminability across the considered SNR range; these performance trends among the benchmark schemes persist throughout the remaining simulations.

\begin{figure}[t]
  \begin{center}
    \includegraphics[width=0.65\columnwidth,keepaspectratio]{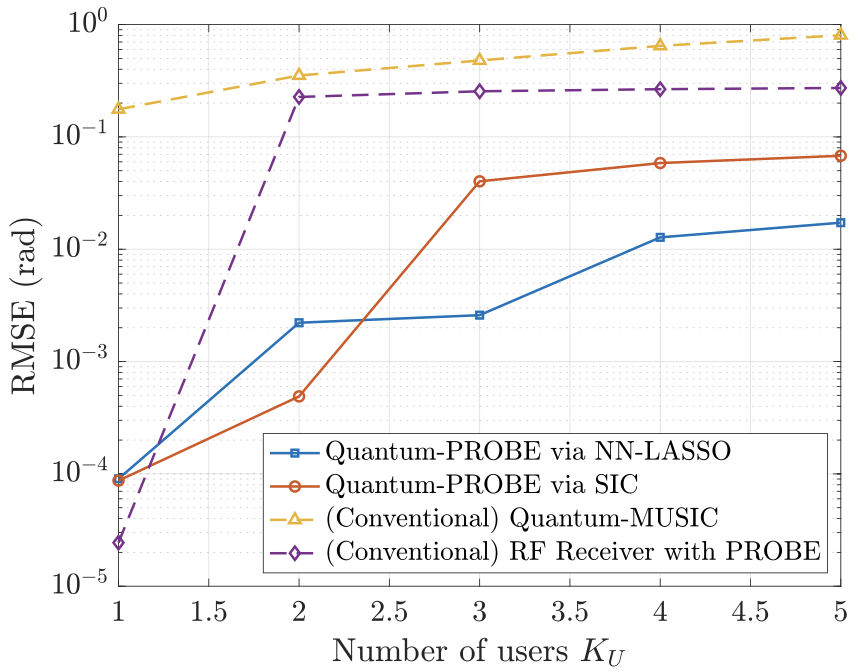}
    \caption{AoA RMSE with respect to $K_{\mathrm{U}}$.}
    \label{fig_ku}
  \end{center}
\end{figure}
Fig.~\ref{fig_ku} illustrates the RMSE performance as a function of $K_{\mathrm{U}}$, where the proposed Quantum-PROBE frameworks consistently outperform all benchmark schemes as the multi-user interference becomes more pronounced. In particular, even at $K_{\mathrm{U}}=5$, Quantum-PROBE via NN-LASSO and SIC maintain RMSEs on the order of $10^{-2}$-$10^{-1}$~rad, whereas Quantum-MUSIC and the RF receiver with PROBE suffer from substantially larger errors, remaining at the $10^{-1}$-$10^{0}$~rad level. In the proposed framework, each additional user contributes a nonnegative, spatially localized power component induced by the RF lens, such that the overall RARE observation naturally follows a superposition model whose structure is preserved even for large $K_{\mathrm U}$, as depicted in Fig.~\ref{fig_hatw} and~\ref{fig_sic}. Hence, the inter-user cross terms remain largely suppressed, preventing error amplification as more users are added.
%The NN-LASSO-based approach explicitly exploits this structure by enforcing sparsity in the angular domain, enabling reliable separation of overlapping AoA components even in densely populated scenarios, whereas the SIC-based implementation progressively cancels the dominant AoA contributions, thereby containing multi-user interference growth with lower computational complexity. By contrast, as $K_{\mathrm U}$ increases, Quantum-MUSIC suffers from cumulative errors in phase recovery and subspace estimation under magnitude-only observations, which rapidly deteriorate the orthogonality between signal and noise subspaces. Similarly, the RF receiver with PROBE becomes increasingly vulnerable to RF-chain noise and distortion as the number of superposed users grows, destabilizing the lens-induced power-focusing patterns and severely degrading AoA identifiability in high-$K_{\mathrm U}$ regimes.

\begin{figure}[t]
  \begin{center}
    \includegraphics[width=0.65\columnwidth,keepaspectratio]{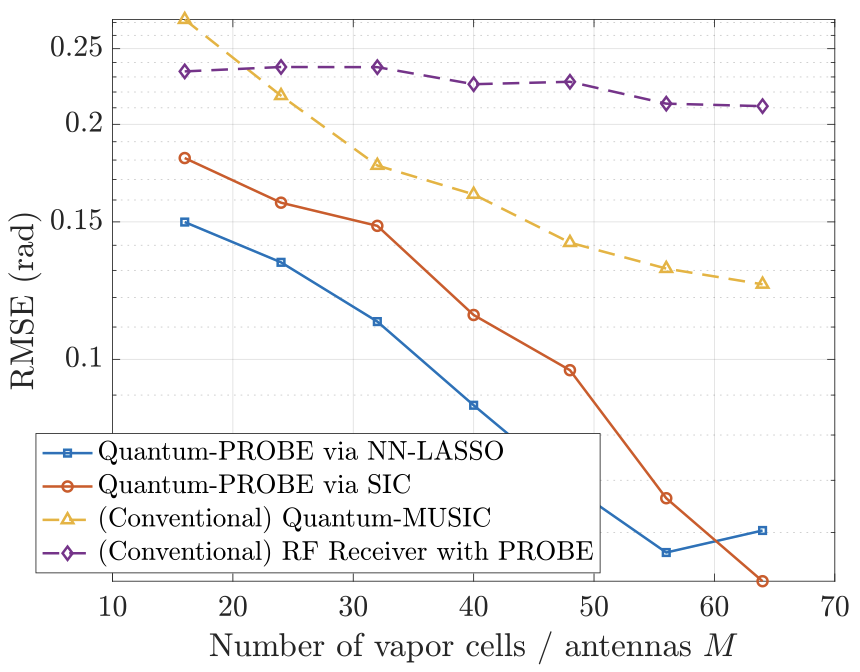}
    \caption{AoA RMSE with respect to the number of vapor cells/antennas $M$.}
    \label{fig_mm}
  \end{center}
\end{figure}

Fig.~\ref{fig_mm} illustrates the RMSE performance as a function of the number of vapor cells/antennas $M$. In particular, at $M=64$, Quantum-PROBE via NN-LASSO and SIC achieve RMSEs of approximately $6\times10^{-2}$ and $5\times10^{-2}$~rad, respectively, which are more than ten times lower than those of Quantum-MUSIC and the RF receiver with PROBE, both of which remain above $10^{-1}$~rad in the same regime. This clear performance gap becomes pronounced as $M$ grows, demonstrating that the proposed Quantum-PROBE frameworks more effectively exploit array scaling, and indicating that they efficiently leverages the increased spatial degree-of-freedom (DoF) and finer sampling resolution of power profile provided by a larger RARE array.
%Herein, the NN-LASSO-based approach benefits from this effect by resolving sparse angular supports with improved spatial resolution, while the SIC-based implementation progressively refines AoA estimates by canceling dominant components, achieving comparable scaling behavior with reduced complexity. In contrast, although Quantum-MUSIC also benefits from increasing $M$ due to improved subspace resolution, its reliance on phase recovery from magnitude-only measurements fundamentally limits the achievable gain, resulting in a slower RMSE decay. Meanwhile, the RF receiver with PROBE shows only marginal improvement as $M$ increases, since RF-noise-induced distortion, which is amplified by $M$, destabilize the lens-induced power profiles, preventing the system from fully capitalizing on the enlarged array. 

Overall, these simulation results demonstrate that the proposed Quantum-PROBE framework offers a favorable accuracy-complexity tradeoff: the NN-LASSO variant provides the highest estimation accuracy at the cost of higher per-iteration overhead, while the SIC-based variant achieves slightly inferior but still competitive estimation performance with significantly reduced computational complexity, making it particularly attractive for large-scale RARE deployments.
\section{Conclusion}
In this paper, we revealed that RF lens-assisted RARE sensing inherently induces a structured nonnegative superposition of AoA-dependent power atoms, whose spatial localization enables reliable multi-user AoA estimation without requiring phase reconstruction. This finding established a fundamental connection between quantum sensing measurements and sparse signal representation, and demonstrates that the lens-induced spatial focusing effect can be systematically exploited to achieve scalable and computationally efficient AoA recovery. The results further highlighted that physics-aware modeling plays a critical role in unlocking interpretable and structured inference mechanisms for quantum-assisted sensing systems. Overall, the proposed Quantum-PROBE provides a unified and scalable approach for multi-user AoA estimation in lens-assisted RARE systems, bridging quantum sensing physics and array signal processing toward 6G wireless sensing applications.

Nevertheless, practical realization of large-scale RARE arrays requires uniform optical excitation across multiple vapor cells and precise optical alignment, where calibration and alignment errors may accumulate and pose a scalability bottleneck due to the complexity of optical distribution and stabilization~\cite{rydalin, rydalin22}. Despite these challenges, the proposed framework, when coupled with the fundamentally ultra-low noise floor enabled by quantum sensing beyond the thermal-noise limit, provides a powerful pathway toward reliable sensing in extremely weak signal regimes, enabling extreme low-SNR and mission-critical applications such as long-range drone detection and surveillance~\cite{lrd, lrd22}, where substantially extended detection ranges beyond conventional RF receivers can be achieved.

%Future research directions include adaptive and data-driven dictionary refinement, extension to dynamic and wideband sensing scenarios, and joint co-design of quantum sensing hardware and signal processing algorithms, paving the way toward robust and high-resolution sensing architectures for emerging 6G applications.
%In this paper, we proposed Quantum-PROBE, a multi-user AoA estimation framework that combines RARE with an RF lens front-end. We first showed that the resulting RARE measurement via BPM can be characterized as a nonnegative superposition of AoA-dependent, non-uniform power profiles, which naturally lends itself to multi-user AoA recovery without phase recovery. Building on this observation, we developed two complementary solutions: a principled NN-LASSO-based formulation that serves as a model-consistent and robust reference approach, and a low-complexity SIC-based algorithm that capitalizes on the strong spatial localization of lens-induced power patterns. Extensive simulations verify that the proposed Quantum-PROBE frameworks consistently outperform existing RARE-based and RF-based benchmarks across diverse operating conditions, and further highlight a clear accuracy-complexity tradeoff that enables selecting an appropriate method for a given sensing environment. Overall, Quantum-PROBE provides a unified and scalable approach for multi-user AoA estimation in lens-assisted RARE systems, bridging quantum sensing physics and array signal processing toward 6G wireless sensing applications.
\bibliographystyle{IEEEtran}
\bibliography{IEEEexample}

\end{document}